\documentclass[a4paper,prb,reprint, twocolumn,superscriptaddress]{revtex4-2}
\usepackage{ulem}
\usepackage{amssymb}
\usepackage{amsmath}
\usepackage{epsfig}
\usepackage{color}
\usepackage{graphics, graphicx}
\usepackage{bbold}
\usepackage{psfrag}
\usepackage{mathcomp}
\usepackage{amsmath}
\usepackage{amssymb}%
\usepackage{mathrsfs}%
\usepackage{subfigure}
\usepackage{verbatim}
\usepackage{xcolor}
\UseRawInputEncoding
\usepackage[colorlinks,citecolor=blue,urlcolor=blue]{hyperref}
\usepackage{mathrsfs}
\usepackage[margin=1in,letterpaper]{geometry} 
\usepackage{ulem}

\begin{document}

	\title{\textbf{Diagnosis of pairing symmetry by vortex and edge spectra in kagome superconductors}}
	\author{Peize Ding}
	\email{peize18@mail.ustc.edu.cn}
	\affiliation{Institute for Theoretical Physics,
		University of W\"{u}rzburg, Am Hubland, D-97074 W\"{u}rzburg, Germany}
	\affiliation{University of Science and Technology of China, Hefei 230026, China}
	\author{Ching Hua Lee}
	\affiliation{Department of Physics, National University of Singapore, Singapore, 117542}
	\author{Xianxin Wu}
	\email{xianxin.wu@fkf.mpg.de}
	\affiliation{Max-Planck-Institut f\"ur Festk\"orperforschung, Heisenbergstrasse 1, D-70569 Stuttgart, Germany}
	\affiliation{CAS Key Laboratory of Theoretical Physics, Institute of Theoretical Physics, Chinese Academy of Sciences, Beijing 100190, China}
	
	\author{Ronny Thomale}
	\email{rthomale@physik.uni-wuerzburg.de}
	\affiliation{Institute for Theoretical Physics,
		University of W\"{u}rzburg, Am Hubland, D-97074 W\"{u}rzburg, Germany}
	\date{\today}
	\begin{abstract}
		Layered kagome metals AV$_3$Sb$_5$ (A=K, Rb, Cs) exhibit diverse correlated electron phenomena. It includes charge density wave formation and superconductivity the pairing symmetry of which, however, is controversial due to contradictory experimental evidence. Through calculations
		based on real-space lattice models at the mean-field level, we investigate the vortex and surface spectra of all competitive pairing propensities suggested for AV$_3$Sb$_5$ from a weak coupling analysis of unconventional superconductivity. Chiral $p$-wave pairing emerges as the only option to host Majorana bound states in the vortex core. We find chiral edge states for both $p$-wave and $d$-wave pairing, along with flat Andreev surface bound states for $f$-wave pairing. Our results expand the fingerprint of superconducting pairing, and thus will contribute to resolving the nature of superconductivity in AV$_3$Sb$_5$.
	\end{abstract}
	\maketitle

	\section{Introduction}
	
	Transition-metal based kagome materials, where the atomic lattice is composed of corner-sharing triangles, offer an exciting platform to explore intriguing correlated and topological phenomena. This includes quantum spin liquids, unconventional superconductivity, Dirac/Weyl semimetals, nematicity, and charge density wave (CDW) order~\cite{spin_liquids, kagome_sc, competing_order, CDW}. Such exotic quantum phenomena are often intimately related to large geometric spin or charge frustration, flat bands and van Hove singularities in the kagome lattice~\cite{unconventional_instability, sub_interference, flat_band1, flat_band2}. Recently, a new family of vanadium-based kagome materials AV$_3$Sb$_5$~\cite{kagomediscovery0} has been discovered featuring $Z_2$ topological invariant in their band structure as well as superconductivity and CDW order at the level of many-body instabilities. The 3d orbitals from the two-dimensional vanadium kagome nets dominate near the Fermi level, indicating strong electronic correlation effects. The maximum $T_c$ is about 2.5 K at ambient pressure while external pressure can enhance $T_c$ up to 8 K with two superconducting domes~\cite{kagome_Tc1, kagome_Tc2, kagome_Tc3, kagome_Tc4, kagome_Tc5, kagome_Tc6, wen2021superconducting}. Moreover, an intriguing CDW order with an inplane 2$\times$2 reconstruction is observed below $T_{CDW}\approx 78-103$ K with time-reversal symmetry breaking and a giant anomalous Hall effect~\cite{kagome_CDW1, kagome_CDW2, kagome_CDW3, kagome_CDW4, chargeorder, lin2021kagome}. Further X-ray scattering measurements suggest that the CDW order is three-dimensional~\cite{x-ray}. For the superconducting gap, penetration depth and nuclear magnetic resonance measurements tend to suggest a nodeless gap~\cite{Duan2021}. However, double superconducting domes under pressure~\cite{Chen2021a}, a significant residual in the thermal conductivity~\cite{zhao2021nodal} and an edge supercurrent in Nb/K$_{1-x}$V$_3$Sb$_5$ indicate unconventional superconductivity with a possible nodal gap. The notion of possible unconventional pairing is further supported by the observation of time-reversal symmetry breaking in the superconducting phase~\cite{mielke2021timereversal}. Today, the pairing symmetry of AV$_3$Sb$_5$ is still intensely debated, where the above synopsis of conflicting experimental evidence emphasizes the need for further means to discriminate between possible pairing states and mechanisms.
	
	Scanning Tunneling Microscopy (STM) measurements at an ultralow temperature unambiguously reveal multiband superconducting features deductible from a V-shaped gap or a U-shaped gap. A unique advantage of STM experiments is to study quasiparticle states and superconducting vortex bound states in real space by applying a magnetic field. For AV$_3$Sb$_5$, a zero-bias conductance peak has been observed in the vortex center on the Cs 2$\times$2 surface and largely exposed Sb surface, which might be a reminiscence of Majorana bound states known to appear in topological superconductors with an odd Bogoliubov Chern number~\cite{vortex_STM1, vortex_STM2}. As the bound states reflect the superconducting gap, the theoretical study of the vortex spectra and topological properties of different pairing states in kagome lattice will be helpful in elucidating the pairing symmetry of AV$_3$Sb$_5$ superconductors through comparing experimental data with theoretical calculations.
	
	In this paper, we investigate the vortex and edge spectra of different pairing states in kagome superconductors by performing calculations based on real-space lattice models. As the AV$_3$Sb$_5$ superconductors are in the vicinity of multiple van Hove singularities, we focus on the vanadium $d$ orbitals and adopt a tight-binding model on kagome lattice whose Fermi surface is consistent with ARPES experiments~\cite{kagome_Tc2}. For the pairing states, we consider all the possible gap functions suggested from a random phase approximation (RPA) analysis of AV$ _3 $Sb$ _5 $~\cite{RPA}. According to our numerical calculations, there are conventional Caroli-de Gennes-Matricon (CdGM) states in the vortex for the $s$-wave state, which would be the pairing of choice assuming conventional pairing in AV$_3$Sb$_5$. From the perspective of unconventional pairing, the $p_x+ip_y$-wave state, characterized by a nontrivial Chern number, hosts Majorana bound states in the vortex core. For chiral $ d $-wave ($ d_{x^2 - y^2} + i d_{xy} $-wave) pairing, two chiral edge states appear on each boundary in the edge spectrum. The vortex spectra are similar to that of the $s$-wave state aside from slight modifications due to the anisotropic gap on the Fermi surface. For two types of $ f $-wave pairings, the local density of states features a broad peak around zero energy. The wavefunction leaks out in the nodal direction, forming a six-pointed star (\cite{f-wave}). Moreover, both $f$-wave pairings host zero-energy Andreev bound states on edges characterized by a nontrivial winding number. We compare our results with available STM measurements. Beyond the context of kagome superconductors AV$ _3 $Sb$ _5 $, our results on vortex and edge spectra promise to be generically applicable to a large class of hexagonal unconventional superconductors. 
	
	The paper is organized as follows. In Sec.~\ref{model_}, we present the three-band tight-binding model for AV$_3$Sb$_5$, the pairing harmonics for different pairing symmetries in the kagome lattice, and the vortex structure we choose in our calculation. In Sec.~\ref{results}, we investigate the vortex and the edge spectra for $ s $, $ p+ip $, $ d+id $, nearest neighbor and next nearest neighbor pairing $ f $-wave pairing. The sharp difference between the vortex spectra and the topological properties of these pairings are emphasized. Finally, in Sec.~\ref{discussion}, we compare our results with available STM experimental data and comment on their ability to determine the pairing symmetry of the kagome superconductors. We also give a complete summary of the main results of this paper.

	\section{Model}\label{model_}
\begin{figure}[tbp]
		\centering
		\includegraphics[width=0.45\textwidth] {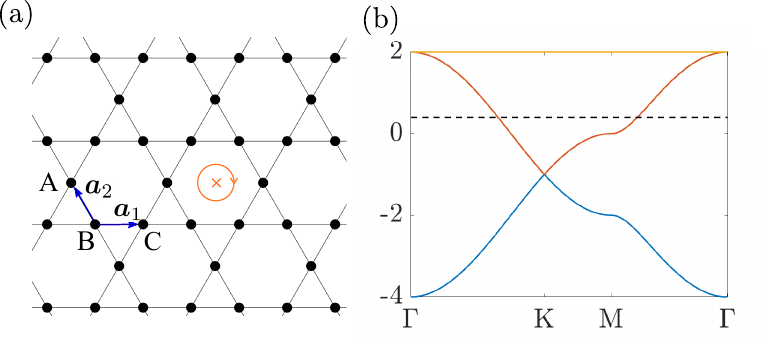}
		\caption{(a) schematic illustration of the kagome lattice and the location of the vortex core. $ \mathbf{a}_{1,2} $ are the sublattice connecting vectors. (b) energy bands of the non-interacting Hamiltonian $ H_{0} $. The chemical potential $ \mu $ is chosen as the experimental motivated value $ 0.4 $, denoted by dashed line in the figure.
		}
		\label{model}
	\end{figure}

	The multi-orbital AV$_3$Sb$_5$ superconductors are in the vicinity of multiple van Hove singularities, derived from vanadium kagome nets. Except the Fermi surfaces from $d$ orbitals, there is an Sb-$p_z$ electron pocket around the Brillouin zone center. To capture the main density of states near the Fermi level and preserve the complexity of multiple van Hove singularities, the minimum model is six-band where two orbitals are located on each sublattice~\cite{RPA}. As RPA calculations suggest dominant intraorbital pairing, we further simplify the model into a three-band one in order to study the general features of pairings in kagome lattice. {The further inclusion of Sb-$p_z$ orbital will only change the results quantitatively due to the weak hybridization between Sb-$p_z$ orbitals and V-$ 3d $ orbitals.}
	The kagome lattice structure is displayed in Fig. \ref{model}(a), where the orange circle denotes a vortex.
	
	The tight-binding (TB) Hamiltonian is $ H_{0} = \sum_{ij} h_{ij} c_{i}^{\dagger} c_{j} = - t \sum_{\langle i, j \rangle} c_{i}^{\dagger} c_{j}  - \mu \sum_{i} c_{i}^{\dagger} c_{i}$ and we further transform it into momentum space, {assuming periodic boundary condition at this moment}
	\begin{eqnarray}
	H_0= -t \sum_{\boldsymbol{k} \alpha \beta} \Phi_{\alpha \beta}(\boldsymbol{k}) c^{\dagger}_{\boldsymbol{k} \beta} c_{\boldsymbol{k} \alpha} - \mu \sum_{\boldsymbol{k} \alpha} c_{\boldsymbol{k} \alpha}^{\dagger} c_{\boldsymbol{k} \alpha}
	,
	\end{eqnarray}
	where the structure factors $ \Phi_{\alpha \beta} $ are given by $ \Phi_{AB} = 1 + e^{-2i \boldsymbol{k}\cdot \boldsymbol{a}_2} $, $ \Phi_{BC} = 1 + e^{-2i \boldsymbol{k}\cdot \boldsymbol{a}_1} $, $ \Phi_{AC} = 1 + e^{-2i \boldsymbol{k}\cdot \boldsymbol{a}_3} $, {$ \Phi_{\beta\alpha} = \Phi_{\alpha\beta}^{\star} $} and $ \boldsymbol{a}_1 = (1, 0) $, $ \boldsymbol{a}_2 = (-\frac{1}{2}, \frac{\sqrt{3}}{2}) $, $ \boldsymbol{a}_3 = \boldsymbol{a}_1 + \boldsymbol{a}_2 $. The dispersion relation of the three-band model is shown in Fig.~\ref{model}(b). We choose the chemical potential $ \mu = 0.4 $, {close to half filling in the kagome lattice}, where the upper van Hove singularity is slightly below the Fermi level and the corresponding Fermi surface is hexagonal, consistent with ARPES experiments and DFT calculations~\cite{kagome_Tc2}. Further including superconducting pairing, the Bogoliubov-de Gennes (BdG) Hamiltonian in real space is given by
	\begin{equation}\label{Hamiltonian}
	\mathcal{H}=\left(\begin{array}{cc}
	\hat{h} & \hat{\Delta} \\
	\hat{\Delta}^{\dagger} & -\hat{h}^{\star}
	\end{array}\right)
	\end{equation}
	where $\hat{h} $ and $ \hat{\Delta} $ are TB and pairing matrices in the lattice site space. The matrix element $ \Delta_{ij} $ equals to $ V_{i-j}\langle c_{j} c_{i} \rangle $ by definition with $ V $ representing the real-space interaction and $ i - j $ represents the relative position separation. As the interaction only depends on the relative position, the center of mass and relative degrees of freedom are decoupled. Therefore we can write $ \Delta_{ij} = \Delta_c \Delta_{i-j} = \Delta_c \sum_{\boldsymbol{k} \alpha \beta} \Delta_{\alpha \beta}(\boldsymbol{k}) e^{i \boldsymbol{k} \cdot(\boldsymbol{r}_{i} - \boldsymbol{r}_{j})} $.
	
	According to the previous RPA study (\cite{RPA}), there are several pairing states with close pairing strengths. For $s, p, d$-wave, the dominant pairing is the lowest harmonic, i.e. pairing between nearest-neighbor (NN) sites, while two types of $ f $-wave pairings corresponding to NN and next NN pairing both might be dominant. What is more, the two-fold degenerate pairings ($p$-wave and $d$-wave) tend to form time-reversal breaking states $p+ip$ or $d+id$ to maximize the superconducting condensation energy. Therefore, in the following calculations, we consider the NN pairing $ p_{x} + i p_{y} $ (chiral $p$-wave), $ s $, $ d_{x^{2} - y^{2}} + i d_{xy} $ (chiral $d$-wave), $ f_{x^{3} - 3 x y^{2}} $-wave, and the next NN $ f_{y^{3} - 3 y x^{2}} $-wave pairing. 
	We define the lattice harmonics between NN sites as
	\begin{equation}\label{key}
	\begin{split}
	&f_{A B}^{1 e}=\cos \left(- \boldsymbol{k}\boldsymbol{a}_{2}\right), \quad f_{A B}^{1 o}=\sin \left(- \boldsymbol{k}\boldsymbol{a}_{2}\right), \\
	&f_{A C}^{1 e}=\cos \left(- \boldsymbol{k}\boldsymbol{a}_{3}\right), \quad f_{A C}^{1 o}=\sin\left(-\boldsymbol{k}\boldsymbol{a}_{3}\right), \\
	&f_{B C}^{1 e}=\cos \left(- \boldsymbol{k}\boldsymbol{a}_{1}\right), \quad f_{B C}^{1 o}=\sin \left(- \boldsymbol{k}\boldsymbol{a}_{1}\right)
	\end{split}
	\end{equation}
	Using the Gell-Man matrices in sublattice space $ \lambda_{1}$-$\lambda_{6} $, we give the pairing harmonics in momentum space
	\begin{equation}\label{key}
	\begin{split}
	\Delta_{1 s}^{S}(\boldsymbol{k}) &=f_{A B}^{1 e} \lambda_{1}+f_{A C}^{1 e} \lambda_{4}+f_{B C}^{1 e} \lambda_{6}\\
	\Delta_{1 d_{x y}}^{S}(\boldsymbol{k}) &=-\sqrt{3} f_{A B}^{1 e} \lambda_{1}+\sqrt{3} f_{A C}^{1 e} \lambda_{4} \\
	\Delta_{1 d_{x^{2}-y^{2}}}^{S}(\boldsymbol{k}) &=f_{A B}^{1 e} \lambda_{1}+f_{A C}^{1 e} \lambda_{4}-2 f_{B C}^{1 e} \lambda_{6}\\
	\Delta_{1 p_{x}}^{T}(\boldsymbol{k})&=f_{A B}^{1 o} \lambda_{1}-f_{A C}^{1 o} \lambda_{4}-2 f_{B C}^{1 o} \lambda_{6}\\
	\Delta_{1 p_{y}}^{T}(\boldsymbol{k})&=\sqrt{3} f_{A B}^{1 o} \lambda_{1}+\sqrt{3} f_{A C}^{1 o} \lambda_{4}, \textsl{}\\
	\Delta_{1 f_{x^{3}-3 x y^{2}}}^{T}(\boldsymbol{k})&=-f_{A B}^{1 o} \lambda_{1}+f_{A C}^{1 o} \lambda_{4}-f_{B C}^{10} \lambda_{6}
	\end{split}
	\end{equation}
	The NNN $ f $-wave pairing is given by
	\begin{equation}\label{key}
	\begin{split}
	\Delta_{2 f_{y^{3}-3 y x^{2}}}^{T}(\boldsymbol{k}) &=
	\sin \left[-\boldsymbol{k}(\boldsymbol{a}_1 + \boldsymbol{a}_2)\right] \lambda_{1}\\
	&+\sin \left[-\boldsymbol{k}(\boldsymbol{a}_2 + \boldsymbol{a}_1)\right]\lambda_{4}\\
	&-\sin \left[-\boldsymbol{k}(\boldsymbol{a}_2 + \boldsymbol{a}_3)\right] \lambda_{6}
	\end{split}
	\end{equation}
	We refer the supplemental material of the paper~\cite{RPA} to the readers for a more complete summary of the pairing harmonics in the kagome lattice.
	
	For type-II superconductors, applying a magnetic field can generate vortices, and a single vortex has two effects on the Hamiltonian. {For clarity, from now on, note that we apply open boundary condition to discuss the vortex related physics.} Firstly, after removing the vector potential by gauge transformation, $ \hat{\Delta} $ acquires a phase winding, i.e. $ \Delta_{ij} \rightarrow \Delta_{ij} e^{-\frac{i}{2}(\theta_{i} + \theta_{j})} $, {where $ \theta_{i, j} $ denotes the polar angle of $ \boldsymbol{r}_{i, j} $ respectively}. Secondly, the pairing potential vanishes at the center of vortex and then increases away from the center and recovers to its asymptotic value outside the vortex core. To simulate this, we choose $ \Delta_{ci} = \Delta\tanh(\frac{r_{i}}{\xi}) $, with $ \Delta $ denoting the pairing strength, $ r_{i} $ the distance from the vortex center, and $ \xi $ being the coherent length, which we choose to be unity (we set the lattice spacing $ a = 1 $ throughout the paper). In our calculations, we put the vortex in the center of the hexagon, maintaining all the crystal symmetries in kagome lattice, as shown in Fig. \ref{model} (a). We diagonalize the real-space Hamiltonian on a parallelogram-shaped sample spanned by the direction of $ \boldsymbol{a}_1 $ and $ \boldsymbol{a}_2 $ with $ N $ unit cells in each dimension.
	
	In order to simulate STM experiments on vortices, we calculate the local density of states per area,
	\begin{equation}\label{key}
	\begin{split}
	{\rm LDOS}(E, \boldsymbol{r}) &= \frac{1}{S} \sum_{n} [|u_{n}(\boldsymbol{r})|^2 \delta(E - E_{n}) \\
	& + |v_{n}(\boldsymbol{r})|^2 \delta(E + E_{n}) ],
	\end{split}
	\end{equation}
	where $ \left(u_{n}(\boldsymbol{r}), v_{n}(\boldsymbol{r})\right)^{T} $ is the eigenvector of the BdG Hamiltonian in real space (Eq.~\ref{Hamiltonian}) with the energy $ E_{n} $, and $ S $ denotes the total area of the sample. To clearly present our results, we sum over the LDOS (divided by two) of the six sites on each hexagon in kagome lattice to form a triangular lattice. We then plot the LDOS on a set of triangular lattice points, starting from the vortex core and going along the $ x $-direction, and labeled by the distance from the vortex center.

	\section{Results}\label{results}
	
	\begin{figure*}[htbp]
		\centering
		\includegraphics[width=1\textwidth] {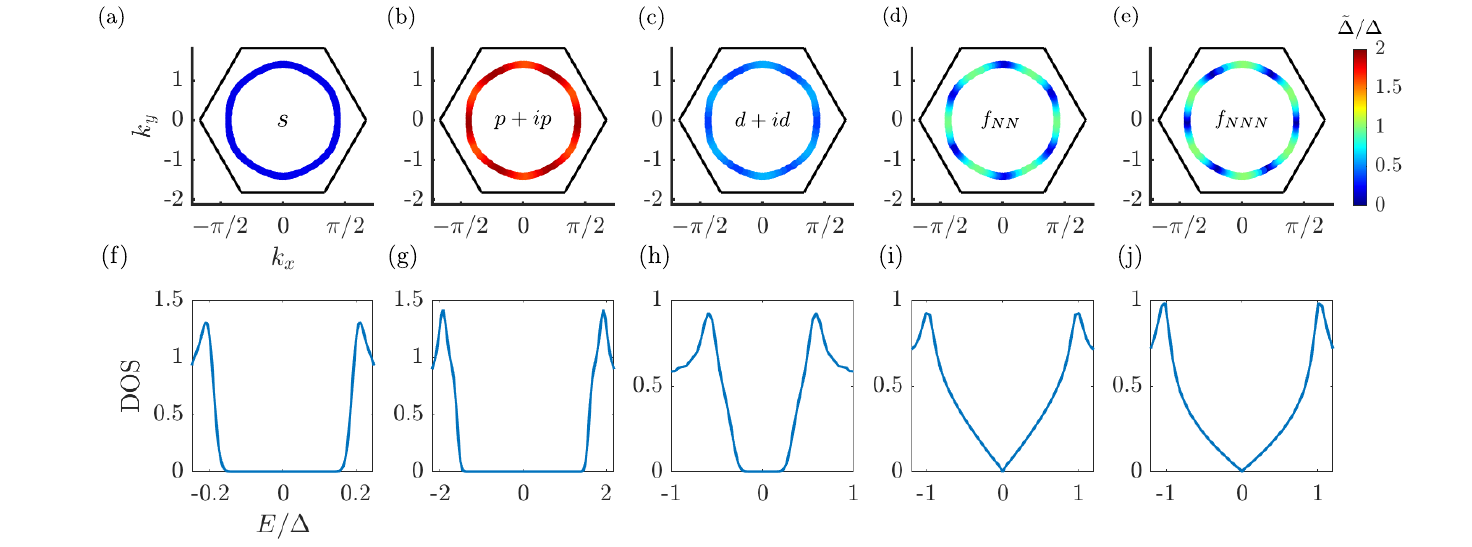}
		\caption{(a)-(e) energy gap evaluated on the Fermi surface $ \tilde{\Delta} $, (f)-(j) DOS per area as a function of energy (lower row) for nearest neighbor s, chiral p, chiral d, f and next nearest neighbor f-wave pairing, respectively. Note that the excitation energy is $ 2\tilde{\Delta} $. $ \Delta = 0.1 $ for chiral $ p $-wave pairing and $ \Delta = 0.2 $ for other pairings.
		}
		\label{gap}
	\end{figure*}
	
	We begin by showing the superconducting gaps on the Fermi surface, and the low-energy density of state (DOS) per for all the pairing states in Fig. \ref{gap}. The chiral $ p $, $ s $, and chiral $ d $-wave states are expected to be fully gapped but the gaps on the Fermi surface for $ s $ and chiral $ d $-wave is relatively small, compared to the NN pairing $\Delta$. This is attributed to the nontrivial sublattice feature of the Fermi surface, in the vicinity of the p-type van Hove singularity~\cite{RPA}. Both NN and NNN $ f $-wave states are gapless owing to the sign change under 60$^\circ$ rotation and their nodal lines are along $\Gamma$-M and $\Gamma$-K, respectively. The $s$-wave gap is isotropic on the Fermi surface while both chiral $p$-wave and $d$-wave gaps exhibit certain anisotropy. In the DOS plots, where the energy is normalized by NN or NNN pairing $\Delta$, the fully gapped states display U-shaped gaps while two $f$-wave states exhibit V-shaped gaps. For chiral $p$-wave and $d$-wave states, there are regions where DOS increases linearly with energy, which is attributed to the gap anisotropy on the Fermi surface. The distinct characteristics of these gaps can be used to distinguish pairing symmetries in kagome lattice when comparing with experiments.   
	
	\begin{figure}[tbp]
		\centering
		\includegraphics[width=0.45\textwidth] {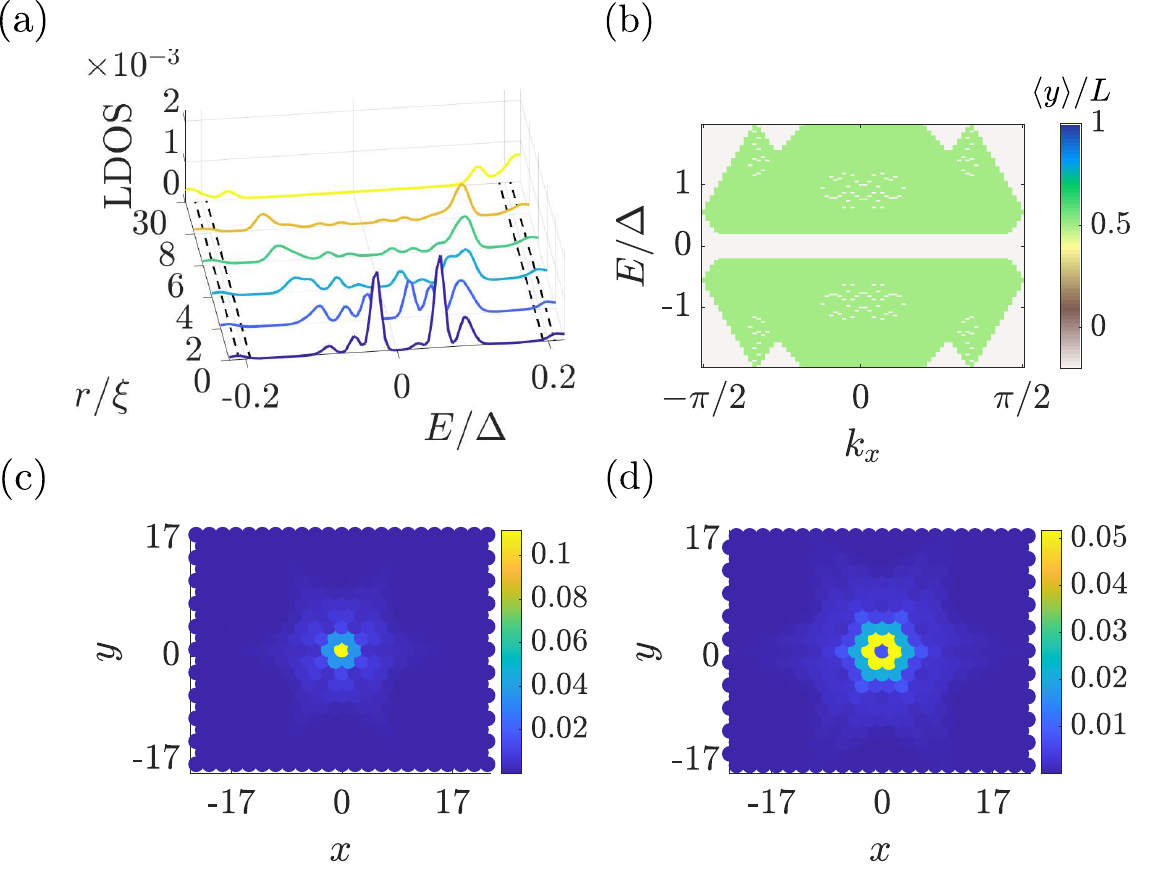}
		\caption{$ s $-wave. (a) Local density of state evaluated at different spatial point as a function of energy. Dashed lines mark the maximum and minimum of the bulk gap. (b) Edge spectrum with 151 unit cells in the open boundary direction. (c), (d) wavefunction of the lowest two vortex core modes with $ n = -1/2, 1/2 $. $ \mu = 0.4 $, $ \Delta = 0.8 $, and $ N = 50 $ are adopted. We find set of CdGM states in the vortex core.
		}
		\label{s-wave}
	\end{figure}

	We then move on to discuss the vortex spectra and topological properties for each pairing state in the kagome lattice. The energies of vortex bound states in conventionally pairing superconductors are $ E_{n} \sim n\frac{\Delta^2}{E_F} $, where $ \Delta $ is the gap on the Fermi surface and $ E_F $ is the Fermi energy~\cite{deGennes, Volovik, BdGmethod}. In realistic calculations, since we fix the coherent length $ \xi $ artificially rather than determining it self-consistently (which does not affect the essential physics of the vortex spectra), the energies of the vortex core states are approximately given by
	\begin{equation}\label{energy}
	E_{n} = n \tilde{\Delta} / (k_F \xi),
	\end{equation}
	where $ \tilde{\Delta} $ is the (uniform) gap on the Fermi surface, $ k_F $ is the Fermi wavevector and $ \xi $ characters the size of the vortex core.  For a $s$-wave pairing, the azimuthal quantum number $n$, which equals to the integral angular momentum quantum number plus $ 1/2 $, is a half integer. Note that we have chosen a small $ \xi $ (=1) which gives rise to decent energy spacing, in order to illustrate the interesting features of the vortex spectra. One has to bear in mind as well that in realistic superconducting materials, the energy spacing between vortex core states is usually much smaller than the gap and may not be resolvable by STM experiments. In kagome lattice, the LDOS for the vortex in the NN $ s $-wave state is shown in Fig.~\ref{s-wave}, where we find the energies of vortex bound states are consistent with the Eq.~\ref{energy}. However, since the gap on the Fermi surface $ \tilde{\Delta} $ is small for $ s $-wave, the energy spacing of the vortex modes is also small, resulting in multiple in-gap bound states. It is also interesting to note that the LDOS peaks with energies slightly lower than the bulk gap evolve with the distance $ r $ from the vortex center (Fig.~\ref{s-wave} (a)). This is attributed to the increasing radius where bound states peak with increasing eigenenergy due to orthogonality of wavefunctions. We show the particle wavefunction of the lowest two vortex core modes with $ n = -1/2, 1/2 $ in Fig.~\ref{s-wave}(c), (d). Both of them are localized at the vortex core. Also, note that they are more extended along certain direction and preserve the $ C_6 $ symmetry, which can be accounted for by the hexagonal Fermi surface. As the $s$-wave pairing is topologically trivial, there is no edge state, as shown in Fig.~\ref{s-wave} (b). The color denotes the expectation value of the $y$ component of the position, i.e. $ \frac{1}{L}\int dy y (|u_n(y)|^2 + |v_n(y)|^2) $, with $L$ being the length along the $ y $-direction in the slab. {The LDOS exhibits particle-hole asymmetry, which is attributed to the vortex bound states (\cite{low_lying}).}
	

	\begin{figure}[tbp]
		\centering
		\includegraphics[width=0.45\textwidth] {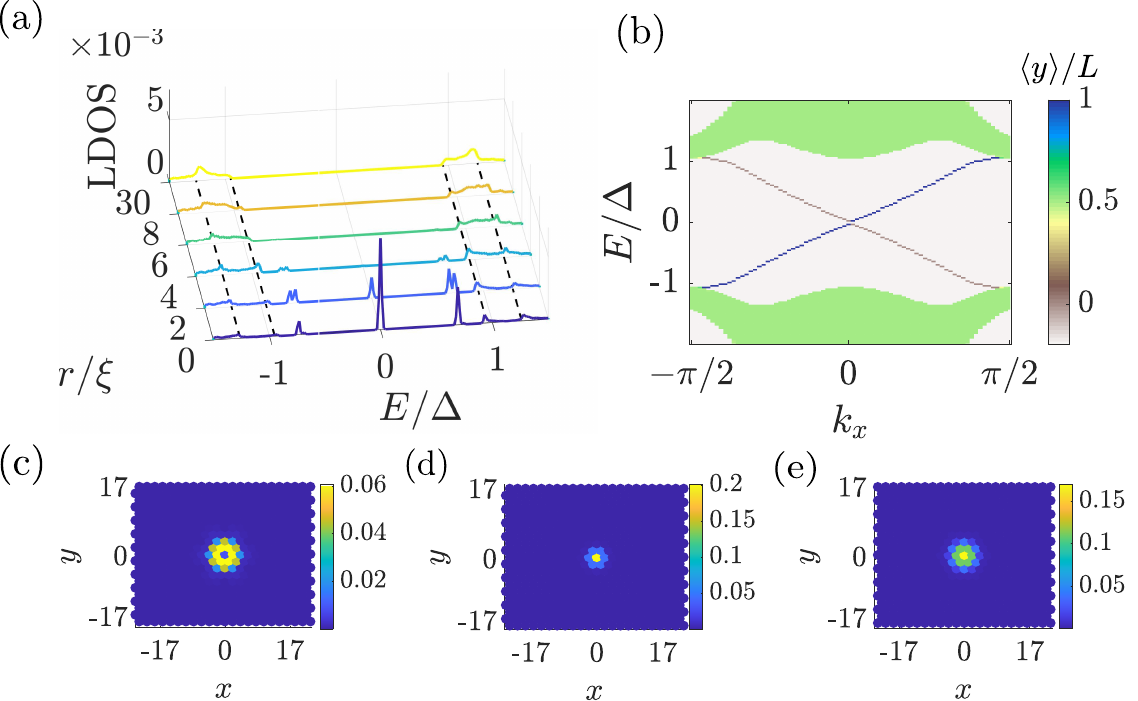}
		\caption{Chiral $ p $-wave. (a) LDOS evaluated at different spatial point as a function of energy. Dashed lines mark the maximum and minimum of the bulk gap. The split of the peaks at $ n = \pm 1 $ result from the anisotropic gap on the Fermi surface. (b) Edge spectrum with 151 unit cells in the open boundary direction. (c)-(e) wavefunction of the three in-gap vortex core modes with angular momentum $ n = -1, 0, 1 $. Note that we choose the mode with lower absolute value of energy to plot from the two states with $ n = 1 $, and, the same for $ n = -1 $. $ \mu = 0.4 $, $ \Delta = 0.5 $ and $ N = 50 $ are adopted. Majorana zero mode is found in the vortex spectrum, and two chiral edge modes appear in the edge spectrum.
		}
		\label{p-wave}
	\end{figure}
	
\begin{figure}[tbp]
		\centering
		\includegraphics[width=0.45\textwidth] {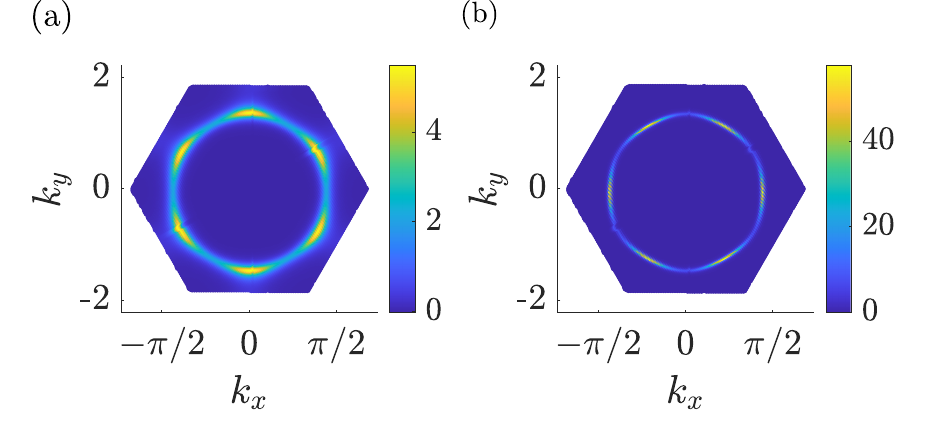}
		\caption{The sum over the Berry curvature of the three negative energy bands of the Bogoliubov-de Gennes Hamiltonian~(Eq.~\ref{Hamiltonian}) for (a) $ p+ip $ and (b) $ d+id $-wave superconductors, with $ \mu = 0.4 $ and $ \Delta = 0.1 $. They give rise to the Chern number $ +1 $ and $ +2 $, respectively.
		}
		\label{Berry}
	\end{figure}

	We further display the LDOS for a vortex in the chiral $ p $-wave state in Fig.~\ref{p-wave} (a). In contrast to $s$-wave pairing, the most prominent feature is the sharp zero-energy peak at the vortex center, which represents the existence of Majorana zero mode (MZM). A detailed analytical illustration showing the existence of MZM in chiral $ p $-wave vortex can be found in \cite{review_p_wave}. Basically, the chiral $ p $-wave superconducting phase is topological with a nontrival Chern number. The vortex core is like an effective boundary, on which topological edge mode lives. The energies of vortex bound states can still be described by Eq.~\ref{energy} but with the angular momentum quantum $n$ being an integer derived from the phase winding of the chiral $p$-wave pairing. In Fig.~\ref{p-wave} (a), one can observe three discrete peaks of LDOS at the vortex core within the bulk gap, whose energies agree with the approximate expression for $ n = 0, \pm 1 $ quite well. In particular, the peaks with $ n = \pm 1 $ split into two, which is ascribed to the anisotropic gap on the Fermi surface. Far away from the vortex center, the pairing potential recovers and the corresponding LDOS is close to bulk DOS, as shown in Fig.~\ref{p-wave} (a) ($ r/\xi = 30 $). The particle wavefunction $ |u_{n}(\boldsymbol{r})|^2 $ converted to triangular lattice (through the same procedure as the LDOS) for the three vortex bound states with $ n = -1, 0, 1 $ are given in Fig.~\ref{p-wave}(c)-(e), respectively. Note that their wavefunctions are dominantly localized at the vortex core and are six-fold symmetrical. The $E_{n=-1}$ state show a six-pointed star pattern around the vortex center, whereas both MZM and $E_{n=1}$ states exhibit solid patterns, in sharp contrast to the $s$-wave case~\cite{low_lying}. We also study the topological properties and the edge spectrum of the chiral $ p $-wave superconductor. In Fig.~\ref{Berry} (a), we plot the sum over the Berry curvature of the three negative energy bands of the Bogoliubov-de Gennes Hamiltonian~(Eq.~\ref{Hamiltonian}). It peaks on the Fermi surface, and, the integration gives the Chern number $ +1 $. We display the edge spectrum with open boundary in $ y $-direction in Fig.~\ref{p-wave}(b). Due to the non-trivial Chern number, there are two symmetric chiral edge modes localized at two boundaries with nearly linear dispersion relation, while the bulk states are extended in the slab.

	\begin{figure}[tbp]
		\centering
		\includegraphics[width=0.45\textwidth] {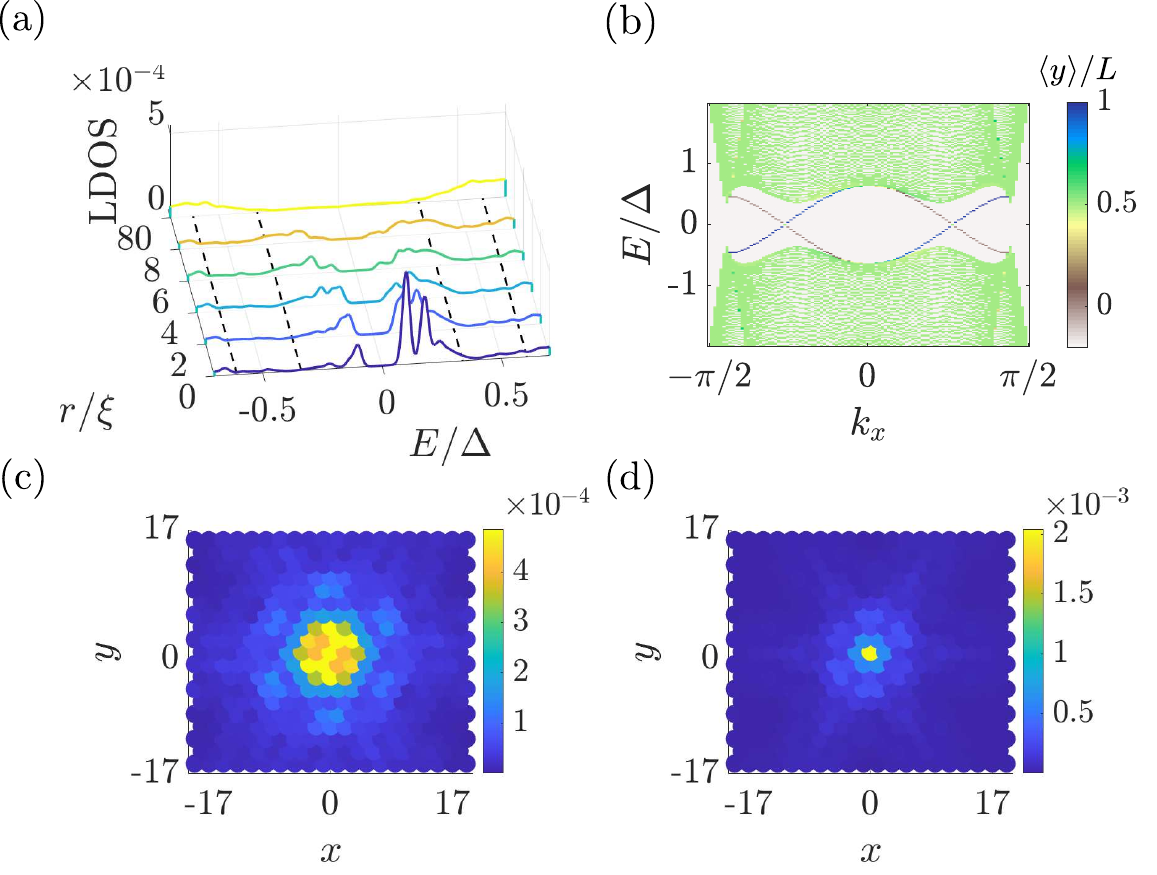}
		\caption{Chiral $ d $-wave. (a) Local density of state evaluated at different spatial point as a function of energy. Dashed lines mark the maximum and minimum of the bulk gap. Two prominent peaks representing vortex core modes are observed. The split of the peaks result from the anisotropic gap on the Fermi surface. (b) Edge spectrum with 221 unit cells in the open boundary direction. (c), (d) wavefunction of the lowest two vortex core modes. The wavefunction extends at directions where the gap is at its minimum. $ \mu = 0.4 $, $ \Delta = 0.2 $ and $ N = 100 $ are adopted. A gap to the lowest core state is found in the vortex spectrum, and four chiral edge modes are observed in the edge spectrum.
		}
		\label{d-wave}
	\end{figure}

	
	Different from the usual gapless $ d_{x^2 - y^2} $-wave pairing where the low energy spectrum at the core is continuous, for the chiral $ d $-wave pairing there exists a gap to the lowest core state \textit{within} bulk gap, in agreement with a previous study in the context of cuprates~\cite{chiral_d}. Two prominent peaks with opposite energies within bulk gap are observed, where the split of each is attributed to the gap anisotropy. The wavefunction of the two lowest vortex core states are shown in Fig.~\ref{d-wave} (c), (d). They are localized at the core while being more extended along the directions of gap minimum. This is qualitatively distinguishable from the vortex core states of $ d_{x^2 - y^2} $-wave pairing superconductors, whose wavefunction leaks out to far from the core in the direction of the nodes. In order to illustrate the topological properties, we also show the Berry curvature summed over the three negative energy bands in Fig.~\ref{Berry} (b), justifying the non-trivial topology and the Chern number $ +2 $ (\cite{d-wave_topology}). The most interesting feature of the edge spectrum for chiral $ d $-wave pairing (Fig.~\ref{d-wave} (b)) is the existence of the four chiral edge modes, two localized at each boundary.
	
	\begin{figure}[tbp]
		\centering
		\includegraphics[width=0.45\textwidth] {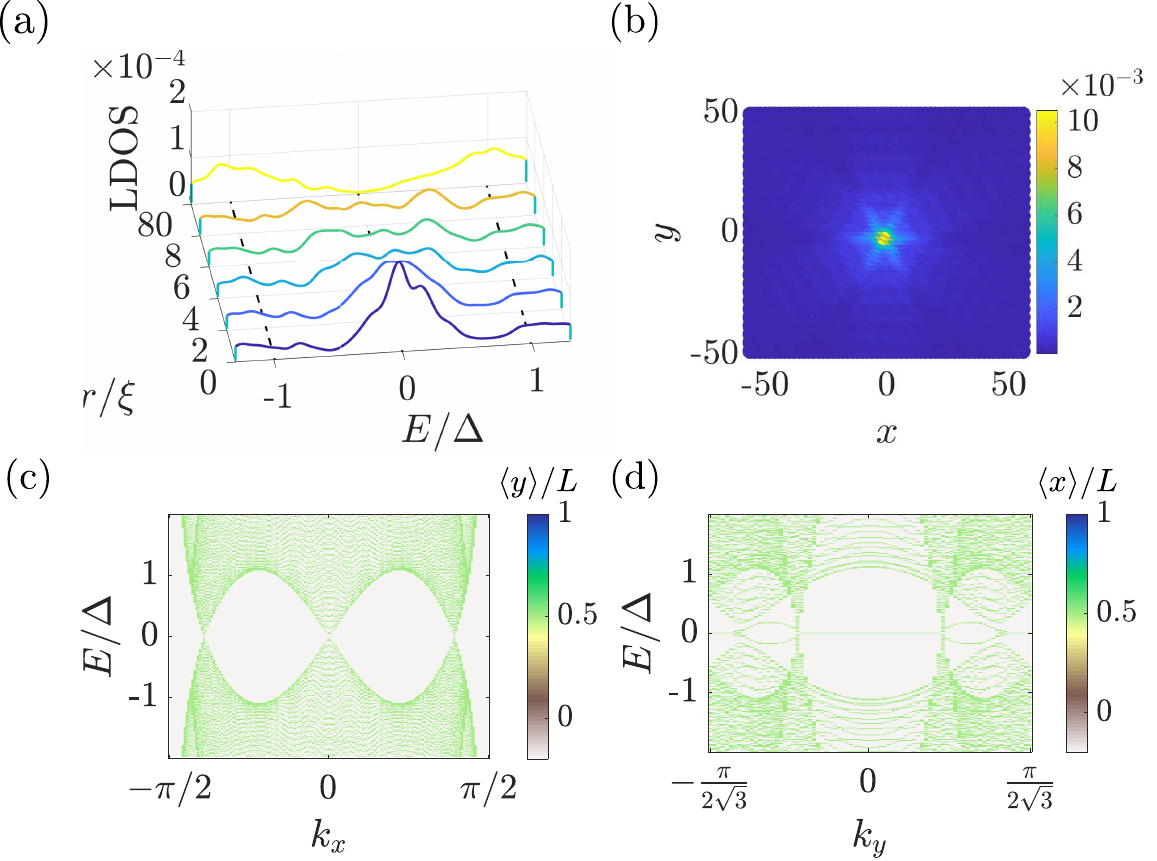}
		\caption{NN $ f $-wave. (a) Local density of state evaluated at different spatial point as a function of energy. Dashed lines mark the maximum and minimum (zero) of the bulk gap. (b) Density of state distribution around the vortex core at $ E = 0 $. We sum over eigenstates with $ |E| < 0.005\Delta $. It extends to out of the core at directions of the nodes. (c) ((d)) Edge spectrum with 331 unit cells with y (x)-direction taken as open boundary condition, respectively. $ \mu = 0.4 $ and $ \Delta = 0.1 $ are adopted. $ N = 100 $ for (a) and $ N = 150 $ for (b). We find a peak around zero energy in the LDOS pattern, and the LDOS distribution around zero energy leaks out of the core along the direction of the nodes. Dispersionless Andreev bound states appear in the edge spectrum with $ x $-direction taken as open boundary condition.
		}
		\label{fNN}
	\end{figure}
	
	\begin{figure}[htbp]
		\centering
		\includegraphics[width=0.45\textwidth] {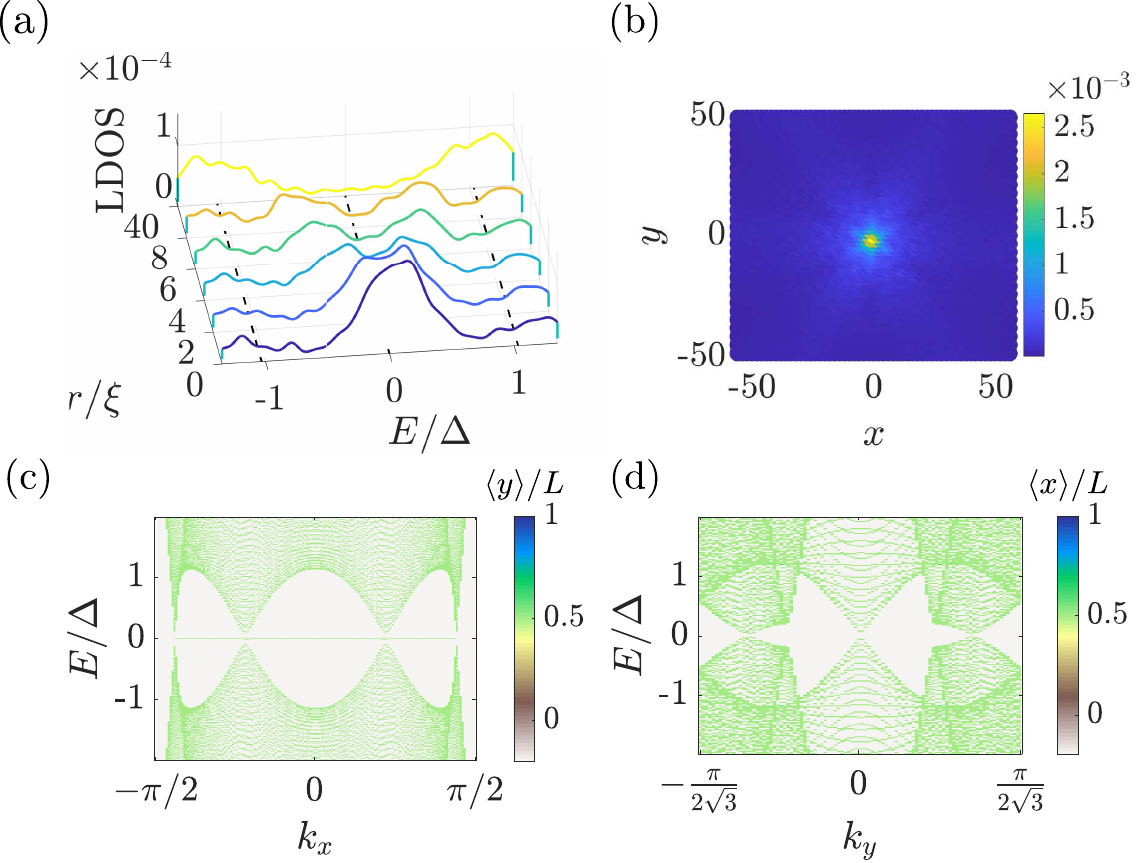}
		\caption{NNN $ f $-wave. (a) Local density of state evaluated at different spatial point as a function of energy.  Dashed lines mark the maximum and minimum (zero) of the bulk gap. (b) Density of state distribution around the vortex core at $ E = 0 $, where we sum over eigenstates with $ |E| < 0.005\Delta $. It extends to out of the core at directions of the nodes. (c) ((d)) Edge spectrum with 331 unit cells with y (x)-direction taken as open boundary condition, respectively. $ \mu = 0.4 $ and $ \Delta = 0.1 $ are adopted. $ N = 100 $ for (a) and $ N = 150 $ for (b). Similar to the NN f-wave, we also observe a peak around zero energy in the LDOS pattern, and the LDOS distribution around zero energy leaks out of the core along the direction of the nodes. Dispersionless Andreev bound states appear in the edge spectrum with $ y $-direction taken as open boundary condition.
		}
		\label{fNNN}
	\end{figure}
	
	We further turn to two gapless $ f $-wave pairings and show the LDOS pattern in Fig.~\ref{fNN} (a) for the NN $ f_{x^3 - 3xy^{2}} $-wave and in Fig.~\ref{fNNN} (a) for next NN pairing $ f_{y^3 - 3yx^{2}} $-wave. There is a broad peak around zero energy in the LDOS and the low energy spectrum at the core becomes continuous in thermodynamic limit. In contrast to fully-gapped states, the $ f $-wave pairing has six nodes on the Fermi surface, which prevents the formation of truly localized vortex bound states. Instead, the vortex core states for nodal pairing states leak out of the core along the nodal directions and they form six-pointed star patterns, as shown in Fig.~\ref{fNN} (b) and Fig.~\ref{fNNN} (b), where we show the LDOS distribution pattern in real space around zero energy around the vortex core for NN and NNN $ f $-wave pairing, respectively.
	
	For time-reversal-invariant superconductors, the topological criterion for the zero-energy Andreev bound states (ABSs) on edges is given by topological winding numbers~\cite{Sato2011,Schnyder2012}. For both $f$-wave pairing states, each node carries a winding number of $+1$ or $-1$. For the slab with an open boundary condition, if the projections of nodes with opposite winding number do not overlap, a zero-energy flat band connecting the projections of nodes will occur. In Fig.~\ref{fNN} (c), (d) (Fig.~\ref{fNNN} (c), (d)), we show the edge spectrum with open boundary in $ y $, $ x $-direction for NN (NNN) $f$-wave pairing, respectively. For the NN $ f_{x^3 - 3xy^{2}} $-wave, dispersionless Andreev bound states (ABS) exist in the edge spectrum with open boundary in $ x $-direction. There are two Andreev modes at zero energy, which are particle-hole partners of each other. The gapped edge states between two ABSs originate from the folding induced hybridization of two edge states with the opposite winding number. In contrast, all projections of nodes with opposite winding number coincide for the open boundary condition along $ y $-direction, inducing the absence of ABSs. Similar reasons hold for the NNN $ f_{y^3 - 3yx^{2}} $-wave state, except that the ABS only appears in the edge spectrum with open boundary in $ y $-direction, for the directions of the nodes are different by a $ \pi / 2 $ rotation from that of $ f_{x^3 - 3xy^{2}} $-wave pairing. The appearance of ABSs will generate a sharp zero-bias peak in the LDOS, which could be observed at corresponding step edges in STM measurements.

	\section{Discussion and Summary} \label{discussion}
	In order to assess the relevance of our theoretical findings, it is necessary to recapitulate the experimental status quo. STM measurements on the vortex spectra of CsV$_3$Sb$_5$ have been performed by two groups~\cite{vortex_STM1, vortex_STM2}. A zero-biased conductance peak at around the vortex core and a clear "$ X $"-type splitting has been observed in the vortex spectra on the Sb surface. Moreover, this peak appears not to split in a large distance when moving away from the vortex center on the Cs 2$\times$2 surface, in superficial agreement with MZM in vortices for chiral $p$-wave pairing. However, the residual DOS at the Fermi energy and a V-shaped gap suggest that there is a nodal gap. The observed six-pointed star shaped vortices further seem to be consistent with $f$-wave pairing. Until today, the data quality is not good enough to pin down the pairing gap and further STM measurements on clean samples at even lower temperature are likely to be necessary to provide clearer evidence for vortex bound states in CsV$_3$Sb$_5$ superconductors.
	
	From our study of vortex and edge spectra for different pairing states in the kagome superconductors, we find the the energies of in-gap vortex bound states to be consistent with $ E_{n} = n \tilde{\Delta} / (k_F \xi) $ for $ s $ and chiral $ p $-wave pairing. The chiral $p$-wave pairing expectedly hosts a MZM in the vortex core. A gap to the lowest localized core state within the bulk gap is observed for chiral $ d $-wave. The topological properties of the chiral $ p $ and chiral $ d $-wave pairings are illustrated through investigating the Berry curvature. For gapless $ f $-wave pairing, we find that the spectrum of vortex core states is continuous in the infinite-size limit, and that the density of state exhibits a broad peak around zero energy. For the edge spectra, the chiral $ p $ and chiral $ d $-wave pairings display chiral edge modes, with dispersionless ABSs for $ f $-wave pairing upon suitably applied boundary conditions is. We expect that the distinct characteristics of vortex spectra and edge modes for different pairings will be helpful in experimentally distinguishing candidate pairing states in AV$_3$Sb$_5$ kagome superconductors.
	
	
	\section*{Acknowledgements}
	
	P.D. is grateful to the Theoretical Physics I group for their hospitality during his visit in University of Würzburg. The work in W\"urzburg  is funded by the Deutsche
	Forschungsgemeinschaft (DFG, German Research Foundation) through
	Project-ID 258499086 - SFB 1170 and through the W\"urzburg-Dresden
	Cluster of Excellence on Complexity and Topology in Quantum Matter - ct.qmat Project-ID 390858490 - EXC 2147.

	\bibliography{kagome_ref}

\begin{thebibliography}{40}%
\makeatletter
\providecommand \@ifxundefined [1]{%
 \@ifx{#1\undefined}
}%
\providecommand \@ifnum [1]{%
 \ifnum #1\expandafter \@firstoftwo
 \else \expandafter \@secondoftwo
 \fi
}%
\providecommand \@ifx [1]{%
 \ifx #1\expandafter \@firstoftwo
 \else \expandafter \@secondoftwo
 \fi
}%
\providecommand \natexlab [1]{#1}%
\providecommand \enquote  [1]{``#1''}%
\providecommand \bibnamefont  [1]{#1}%
\providecommand \bibfnamefont [1]{#1}%
\providecommand \citenamefont [1]{#1}%
\providecommand \href@noop [0]{\@secondoftwo}%
\providecommand \href [0]{\begingroup \@sanitize@url \@href}%
\providecommand \@href[1]{\@@startlink{#1}\@@href}%
\providecommand \@@href[1]{\endgroup#1\@@endlink}%
\providecommand \@sanitize@url [0]{\catcode `\\12\catcode `\$12\catcode
  `\&12\catcode `\#12\catcode `\^12\catcode `\_12\catcode `\%12\relax}%
\providecommand \@@startlink[1]{}%
\providecommand \@@endlink[0]{}%
\providecommand \url  [0]{\begingroup\@sanitize@url \@url }%
\providecommand \@url [1]{\endgroup\@href {#1}{\urlprefix }}%
\providecommand \urlprefix  [0]{URL }%
\providecommand \Eprint [0]{\href }%
\providecommand \doibase [0]{https://doi.org/}%
\providecommand \selectlanguage [0]{\@gobble}%
\providecommand \bibinfo  [0]{\@secondoftwo}%
\providecommand \bibfield  [0]{\@secondoftwo}%
\providecommand \translation [1]{[#1]}%
\providecommand \BibitemOpen [0]{}%
\providecommand \bibitemStop [0]{}%
\providecommand \bibitemNoStop [0]{.\EOS\space}%
\providecommand \EOS [0]{\spacefactor3000\relax}%
\providecommand \BibitemShut  [1]{\csname bibitem#1\endcsname}%
\let\auto@bib@innerbib\@empty
\bibitem [{\citenamefont {Norman}(2016)}]{spin_liquids}%
  \BibitemOpen
  \bibfield  {author} {\bibinfo {author} {\bibfnamefont {M.~R.}\ \bibnamefont
  {Norman}},\ }\bibfield  {title} {\bibinfo {title} {Colloquium:
  Herbertsmithite and the search for the quantum spin liquid},\ }\href
  {https://doi.org/10.1103/RevModPhys.88.041002} {\bibfield  {journal}
  {\bibinfo  {journal} {Reviews of Modern Physics}\ }\textbf {\bibinfo {volume}
  {88}},\ \bibinfo {pages} {041002} (\bibinfo {year} {2016})}\BibitemShut
  {NoStop}%
\bibitem [{\citenamefont {Yu}\ and\ \citenamefont {Li}(2012)}]{kagome_sc}%
  \BibitemOpen
  \bibfield  {author} {\bibinfo {author} {\bibfnamefont {S.-L.}\ \bibnamefont
  {Yu}}\ and\ \bibinfo {author} {\bibfnamefont {J.-X.}\ \bibnamefont {Li}},\
  }\bibfield  {title} {\bibinfo {title} {Chiral superconducting phase and
  chiral spin-density-wave phase in a hubbard model on the kagome lattice},\
  }\href {https://doi.org/10.1103/PhysRevB.85.144402} {\bibfield  {journal}
  {\bibinfo  {journal} {Physical Review B}\ }\textbf {\bibinfo {volume} {85}},\
  \bibinfo {pages} {144402} (\bibinfo {year} {2012})}\BibitemShut {NoStop}%
\bibitem [{\citenamefont {Wang}\ \emph {et~al.}(2013)\citenamefont {Wang},
  \citenamefont {Li}, \citenamefont {Xiang},\ and\ \citenamefont
  {Wang}}]{competing_order}%
  \BibitemOpen
  \bibfield  {author} {\bibinfo {author} {\bibfnamefont {W.-S.}\ \bibnamefont
  {Wang}}, \bibinfo {author} {\bibfnamefont {Z.-Z.}\ \bibnamefont {Li}},
  \bibinfo {author} {\bibfnamefont {Y.-Y.}\ \bibnamefont {Xiang}},\ and\
  \bibinfo {author} {\bibfnamefont {Q.-H.}\ \bibnamefont {Wang}},\ }\bibfield
  {title} {\bibinfo {title} {Competing electronic orders on kagome lattices at
  van hove filling},\ }\href {https://doi.org/10.1103/PhysRevB.87.115135}
  {\bibfield  {journal} {\bibinfo  {journal} {Physical Review B}\ }\textbf
  {\bibinfo {volume} {87}},\ \bibinfo {pages} {115135} (\bibinfo {year}
  {2013})}\BibitemShut {NoStop}%
\bibitem [{\citenamefont {Nayak}(2000)}]{CDW}%
  \BibitemOpen
  \bibfield  {author} {\bibinfo {author} {\bibfnamefont {C.}~\bibnamefont
  {Nayak}},\ }\bibfield  {title} {\bibinfo {title} {Density-wave states of
  nonzero angular momentum},\ }\href {https://doi.org/10.1103/PhysRevB.62.4880}
  {\bibfield  {journal} {\bibinfo  {journal} {Physical Review B}\ }\textbf
  {\bibinfo {volume} {62}},\ \bibinfo {pages} {4880} (\bibinfo {year}
  {2000})}\BibitemShut {NoStop}%
\bibitem [{\citenamefont {Kiesel}\ \emph {et~al.}(2013)\citenamefont {Kiesel},
  \citenamefont {Platt},\ and\ \citenamefont
  {Thomale}}]{unconventional_instability}%
  \BibitemOpen
  \bibfield  {author} {\bibinfo {author} {\bibfnamefont {M.~L.}\ \bibnamefont
  {Kiesel}}, \bibinfo {author} {\bibfnamefont {C.}~\bibnamefont {Platt}},\ and\
  \bibinfo {author} {\bibfnamefont {R.}~\bibnamefont {Thomale}},\ }\bibfield
  {title} {\bibinfo {title} {Unconventional fermi surface instabilities in the
  kagome hubbard model},\ }\href
  {https://doi.org/10.1103/PhysRevLett.110.126405} {\bibfield  {journal}
  {\bibinfo  {journal} {Physical Review Letters}\ }\textbf {\bibinfo {volume}
  {110}},\ \bibinfo {pages} {126405} (\bibinfo {year} {2013})}\BibitemShut
  {NoStop}%
\bibitem [{\citenamefont {Kiesel}\ and\ \citenamefont
  {Thomale}(2012)}]{sub_interference}%
  \BibitemOpen
  \bibfield  {author} {\bibinfo {author} {\bibfnamefont {M.~L.}\ \bibnamefont
  {Kiesel}}\ and\ \bibinfo {author} {\bibfnamefont {R.}~\bibnamefont
  {Thomale}},\ }\bibfield  {title} {\bibinfo {title} {Sublattice interference
  in the kagome hubbard model},\ }\href
  {https://doi.org/10.1103/PhysRevB.86.121105} {\bibfield  {journal} {\bibinfo
  {journal} {Physical Review B}\ }\textbf {\bibinfo {volume} {86}},\ \bibinfo
  {pages} {121105} (\bibinfo {year} {2012})}\BibitemShut {NoStop}%
\bibitem [{\citenamefont {Parameswaran}\ \emph {et~al.}(2013)\citenamefont
  {Parameswaran}, \citenamefont {Roy},\ and\ \citenamefont
  {Sondhi}}]{flat_band1}%
  \BibitemOpen
  \bibfield  {author} {\bibinfo {author} {\bibfnamefont {S.~A.}\ \bibnamefont
  {Parameswaran}}, \bibinfo {author} {\bibfnamefont {R.}~\bibnamefont {Roy}},\
  and\ \bibinfo {author} {\bibfnamefont {S.~L.}\ \bibnamefont {Sondhi}},\
  }\bibfield  {title} {\bibinfo {title} {Fractional quantum hall physics in
  topological flat bands},\ }\href
  {https://doi.org/https://doi.org/10.1016/j.crhy.2013.04.003} {\bibfield
  {journal} {\bibinfo  {journal} {Comptes Rendus Physique}\ }\textbf {\bibinfo
  {volume} {14}},\ \bibinfo {pages} {816} (\bibinfo {year} {2013})}\BibitemShut
  {NoStop}%
\bibitem [{\citenamefont {Lee}\ \emph {et~al.}(2016)\citenamefont {Lee},
  \citenamefont {Arovas},\ and\ \citenamefont {Thomale}}]{flat_band2}%
  \BibitemOpen
  \bibfield  {author} {\bibinfo {author} {\bibfnamefont {C.~H.}\ \bibnamefont
  {Lee}}, \bibinfo {author} {\bibfnamefont {D.~P.}\ \bibnamefont {Arovas}},\
  and\ \bibinfo {author} {\bibfnamefont {R.}~\bibnamefont {Thomale}},\
  }\bibfield  {title} {\bibinfo {title} {Band flatness optimization through
  complex analysis},\ }\href {https://doi.org/10.1103/PhysRevB.93.155155}
  {\bibfield  {journal} {\bibinfo  {journal} {Physical Review B}\ }\textbf
  {\bibinfo {volume} {93}},\ \bibinfo {pages} {155155} (\bibinfo {year}
  {2016})}\BibitemShut {NoStop}%
\bibitem [{\citenamefont {Ortiz}\ \emph {et~al.}(2019)\citenamefont {Ortiz},
  \citenamefont {Gomes}, \citenamefont {Morey}, \citenamefont {Winiarski},
  \citenamefont {Bordelon}, \citenamefont {Mangum}, \citenamefont {Oswald},
  \citenamefont {Rodriguez-Rivera}, \citenamefont {Neilson}, \citenamefont
  {Wilson}, \citenamefont {Ertekin}, \citenamefont {McQueen},\ and\
  \citenamefont {Toberer}}]{kagomediscovery0}%
  \BibitemOpen
  \bibfield  {author} {\bibinfo {author} {\bibfnamefont {B.~R.}\ \bibnamefont
  {Ortiz}}, \bibinfo {author} {\bibfnamefont {L.~C.}\ \bibnamefont {Gomes}},
  \bibinfo {author} {\bibfnamefont {J.~R.}\ \bibnamefont {Morey}}, \bibinfo
  {author} {\bibfnamefont {M.}~\bibnamefont {Winiarski}}, \bibinfo {author}
  {\bibfnamefont {M.}~\bibnamefont {Bordelon}}, \bibinfo {author}
  {\bibfnamefont {J.~S.}\ \bibnamefont {Mangum}}, \bibinfo {author}
  {\bibfnamefont {I.~W.~H.}\ \bibnamefont {Oswald}}, \bibinfo {author}
  {\bibfnamefont {J.~A.}\ \bibnamefont {Rodriguez-Rivera}}, \bibinfo {author}
  {\bibfnamefont {J.~R.}\ \bibnamefont {Neilson}}, \bibinfo {author}
  {\bibfnamefont {S.~D.}\ \bibnamefont {Wilson}}, \bibinfo {author}
  {\bibfnamefont {E.}~\bibnamefont {Ertekin}}, \bibinfo {author} {\bibfnamefont
  {T.~M.}\ \bibnamefont {McQueen}},\ and\ \bibinfo {author} {\bibfnamefont
  {E.~S.}\ \bibnamefont {Toberer}},\ }\bibfield  {title} {\bibinfo {title} {New
  kagome prototype materials: discovery of
  ${\mathrm{kv}}_{3}{\mathrm{sb}}_{5},{\mathrm{rbv}}_{3}{\mathrm{sb}}_{5}$, and
  ${\mathrm{csv}}_{3}{\mathrm{sb}}_{5}$},\ }\href
  {https://doi.org/10.1103/PhysRevMaterials.3.094407} {\bibfield  {journal}
  {\bibinfo  {journal} {Physical Review Materials}\ }\textbf {\bibinfo {volume}
  {3}},\ \bibinfo {pages} {094407} (\bibinfo {year} {2019})}\BibitemShut
  {NoStop}%
\bibitem [{\citenamefont {Yin}\ \emph {et~al.}(2021)\citenamefont {Yin},
  \citenamefont {Tu}, \citenamefont {Gong}, \citenamefont {Fu}, \citenamefont
  {Yan},\ and\ \citenamefont {Lei}}]{kagome_Tc1}%
  \BibitemOpen
  \bibfield  {author} {\bibinfo {author} {\bibfnamefont {Q.}~\bibnamefont
  {Yin}}, \bibinfo {author} {\bibfnamefont {Z.}~\bibnamefont {Tu}}, \bibinfo
  {author} {\bibfnamefont {C.}~\bibnamefont {Gong}}, \bibinfo {author}
  {\bibfnamefont {Y.}~\bibnamefont {Fu}}, \bibinfo {author} {\bibfnamefont
  {S.}~\bibnamefont {Yan}},\ and\ \bibinfo {author} {\bibfnamefont
  {H.}~\bibnamefont {Lei}},\ }\bibfield  {title} {\bibinfo {title}
  {Superconductivity and normal-state properties of kagome metal rbv3sb5 single
  crystals},\ }\href {https://doi.org/10.1088/0256-307x/38/3/037403} {\bibfield
   {journal} {\bibinfo  {journal} {Chinese Physics Letters}\ }\textbf {\bibinfo
  {volume} {38}},\ \bibinfo {pages} {037403} (\bibinfo {year}
  {2021})}\BibitemShut {NoStop}%
\bibitem [{\citenamefont {Ortiz}\ \emph {et~al.}(2020)\citenamefont {Ortiz},
  \citenamefont {Teicher}, \citenamefont {Hu}, \citenamefont {Zuo},
  \citenamefont {Sarte}, \citenamefont {Schueller}, \citenamefont {Abeykoon},
  \citenamefont {Krogstad}, \citenamefont {Rosenkranz}, \citenamefont {Osborn},
  \citenamefont {Seshadri}, \citenamefont {Balents}, \citenamefont {He},\ and\
  \citenamefont {Wilson}}]{kagome_Tc2}%
  \BibitemOpen
  \bibfield  {author} {\bibinfo {author} {\bibfnamefont {B.~R.}\ \bibnamefont
  {Ortiz}}, \bibinfo {author} {\bibfnamefont {S.~M.~L.}\ \bibnamefont
  {Teicher}}, \bibinfo {author} {\bibfnamefont {Y.}~\bibnamefont {Hu}},
  \bibinfo {author} {\bibfnamefont {J.~L.}\ \bibnamefont {Zuo}}, \bibinfo
  {author} {\bibfnamefont {P.~M.}\ \bibnamefont {Sarte}}, \bibinfo {author}
  {\bibfnamefont {E.~C.}\ \bibnamefont {Schueller}}, \bibinfo {author}
  {\bibfnamefont {A.~â.~M.}\ \bibnamefont {Abeykoon}}, \bibinfo {author}
  {\bibfnamefont {M.~J.}\ \bibnamefont {Krogstad}}, \bibinfo {author}
  {\bibfnamefont {S.}~\bibnamefont {Rosenkranz}}, \bibinfo {author}
  {\bibfnamefont {R.}~\bibnamefont {Osborn}}, \bibinfo {author} {\bibfnamefont
  {R.}~\bibnamefont {Seshadri}}, \bibinfo {author} {\bibfnamefont
  {L.}~\bibnamefont {Balents}}, \bibinfo {author} {\bibfnamefont
  {J.}~\bibnamefont {He}},\ and\ \bibinfo {author} {\bibfnamefont {S.~D.}\
  \bibnamefont {Wilson}},\ }\bibfield  {title} {\bibinfo {title}
  {$\mathrm{Cs}{\mathrm{v}}_{3}{\mathrm{sb}}_{5}$: A ${\mathbb{z}}_{2}$
  topological kagome metal with a superconducting ground state},\ }\href
  {https://doi.org/10.1103/PhysRevLett.125.247002} {\bibfield  {journal}
  {\bibinfo  {journal} {Physical Review Letters}\ }\textbf {\bibinfo {volume}
  {125}},\ \bibinfo {pages} {247002} (\bibinfo {year} {2020})}\BibitemShut
  {NoStop}%
\bibitem [{\citenamefont {Zhao}\ \emph
  {et~al.}(2021{\natexlab{a}})\citenamefont {Zhao}, \citenamefont {Li},
  \citenamefont {Ortiz}, \citenamefont {Teicher}, \citenamefont {Park},
  \citenamefont {Ye}, \citenamefont {Wang}, \citenamefont {Balents},
  \citenamefont {Wilson},\ and\ \citenamefont {Zeljkovic}}]{kagome_Tc3}%
  \BibitemOpen
  \bibfield  {author} {\bibinfo {author} {\bibfnamefont {H.}~\bibnamefont
  {Zhao}}, \bibinfo {author} {\bibfnamefont {H.}~\bibnamefont {Li}}, \bibinfo
  {author} {\bibfnamefont {B.~R.}\ \bibnamefont {Ortiz}}, \bibinfo {author}
  {\bibfnamefont {S.~M.~L.}\ \bibnamefont {Teicher}}, \bibinfo {author}
  {\bibfnamefont {T.}~\bibnamefont {Park}}, \bibinfo {author} {\bibfnamefont
  {M.}~\bibnamefont {Ye}}, \bibinfo {author} {\bibfnamefont {Z.}~\bibnamefont
  {Wang}}, \bibinfo {author} {\bibfnamefont {L.}~\bibnamefont {Balents}},
  \bibinfo {author} {\bibfnamefont {S.~D.}\ \bibnamefont {Wilson}},\ and\
  \bibinfo {author} {\bibfnamefont {I.}~\bibnamefont {Zeljkovic}},\ }\bibfield
  {title} {\bibinfo {title} {Cascade of correlated electron states in the
  kagome superconductor csv3sb5},\ }\href
  {https://doi.org/10.1038/s41586-021-03946-w} {\bibfield  {journal} {\bibinfo
  {journal} {Nature}\ }\textbf {\bibinfo {volume} {599}},\ \bibinfo {pages}
  {216} (\bibinfo {year} {2021}{\natexlab{a}})}\BibitemShut {NoStop}%
\bibitem [{\citenamefont {Chen}\ \emph
  {et~al.}(2021{\natexlab{a}})\citenamefont {Chen}, \citenamefont {Wang},
  \citenamefont {Yin}, \citenamefont {Gu}, \citenamefont {Jiang}, \citenamefont
  {Tu}, \citenamefont {Gong}, \citenamefont {Uwatoko}, \citenamefont {Sun},
  \citenamefont {Lei}, \citenamefont {Hu},\ and\ \citenamefont
  {Cheng}}]{kagome_Tc4}%
  \BibitemOpen
  \bibfield  {author} {\bibinfo {author} {\bibfnamefont {K.~Y.}\ \bibnamefont
  {Chen}}, \bibinfo {author} {\bibfnamefont {N.}~\bibnamefont {Wang}}, \bibinfo
  {author} {\bibfnamefont {Q.}~\bibnamefont {Yin}}, \bibinfo {author}
  {\bibfnamefont {Y.}~\bibnamefont {Gu}}, \bibinfo {author} {\bibfnamefont
  {K.}~\bibnamefont {Jiang}}, \bibinfo {author} {\bibfnamefont
  {Z.}~\bibnamefont {Tu}}, \bibinfo {author} {\bibfnamefont {C.~S.}\
  \bibnamefont {Gong}}, \bibinfo {author} {\bibfnamefont {Y.}~\bibnamefont
  {Uwatoko}}, \bibinfo {author} {\bibfnamefont {J.}~\bibnamefont {Sun}},
  \bibinfo {author} {\bibfnamefont {H.}~\bibnamefont {Lei}}, \bibinfo {author}
  {\bibfnamefont {J.}~\bibnamefont {Hu}},\ and\ \bibinfo {author}
  {\bibfnamefont {J.~G.}\ \bibnamefont {Cheng}},\ }\bibfield  {title} {\bibinfo
  {title} {Double superconducting dome and triple enhancement of ${T}_{c}$ in
  the kagome superconductor ${\mathrm{csv}}_{3}{\mathrm{sb}}_{5}$ under high
  pressure},\ }\href {https://doi.org/10.1103/PhysRevLett.126.247001}
  {\bibfield  {journal} {\bibinfo  {journal} {Physical Review Letters}\
  }\textbf {\bibinfo {volume} {126}},\ \bibinfo {pages} {247001} (\bibinfo
  {year} {2021}{\natexlab{a}})}\BibitemShut {NoStop}%
\bibitem [{\citenamefont {Zhang}\ \emph {et~al.}(2021)\citenamefont {Zhang},
  \citenamefont {Chen}, \citenamefont {Zhou}, \citenamefont {Yuan},
  \citenamefont {Wang}, \citenamefont {Wang}, \citenamefont {Yang},
  \citenamefont {An}, \citenamefont {Zhang}, \citenamefont {Zhu}, \citenamefont
  {Zhou}, \citenamefont {Chen}, \citenamefont {Zhou},\ and\ \citenamefont
  {Yang}}]{kagome_Tc5}%
  \BibitemOpen
  \bibfield  {author} {\bibinfo {author} {\bibfnamefont {Z.}~\bibnamefont
  {Zhang}}, \bibinfo {author} {\bibfnamefont {Z.}~\bibnamefont {Chen}},
  \bibinfo {author} {\bibfnamefont {Y.}~\bibnamefont {Zhou}}, \bibinfo {author}
  {\bibfnamefont {Y.}~\bibnamefont {Yuan}}, \bibinfo {author} {\bibfnamefont
  {S.}~\bibnamefont {Wang}}, \bibinfo {author} {\bibfnamefont {J.}~\bibnamefont
  {Wang}}, \bibinfo {author} {\bibfnamefont {H.}~\bibnamefont {Yang}}, \bibinfo
  {author} {\bibfnamefont {C.}~\bibnamefont {An}}, \bibinfo {author}
  {\bibfnamefont {L.}~\bibnamefont {Zhang}}, \bibinfo {author} {\bibfnamefont
  {X.}~\bibnamefont {Zhu}}, \bibinfo {author} {\bibfnamefont {Y.}~\bibnamefont
  {Zhou}}, \bibinfo {author} {\bibfnamefont {X.}~\bibnamefont {Chen}}, \bibinfo
  {author} {\bibfnamefont {J.}~\bibnamefont {Zhou}},\ and\ \bibinfo {author}
  {\bibfnamefont {Z.}~\bibnamefont {Yang}},\ }\bibfield  {title} {\bibinfo
  {title} {Pressure-induced reemergence of superconductivity in the topological
  kagome metal $\mathrm{Cs}{\mathrm{v}}_{3}{\mathrm{sb}}_{5}$},\ }\href
  {https://doi.org/10.1103/PhysRevB.103.224513} {\bibfield  {journal} {\bibinfo
   {journal} {Physical Review B}\ }\textbf {\bibinfo {volume} {103}},\ \bibinfo
  {pages} {224513} (\bibinfo {year} {2021})}\BibitemShut {NoStop}%
\bibitem [{\citenamefont {Chen}\ \emph
  {et~al.}(2021{\natexlab{b}})\citenamefont {Chen}, \citenamefont {Zhan},
  \citenamefont {Wang}, \citenamefont {Deng}, \citenamefont {Liu},
  \citenamefont {Chen}, \citenamefont {Guo},\ and\ \citenamefont
  {Chen}}]{kagome_Tc6}%
  \BibitemOpen
  \bibfield  {author} {\bibinfo {author} {\bibfnamefont {X.}~\bibnamefont
  {Chen}}, \bibinfo {author} {\bibfnamefont {X.}~\bibnamefont {Zhan}}, \bibinfo
  {author} {\bibfnamefont {X.}~\bibnamefont {Wang}}, \bibinfo {author}
  {\bibfnamefont {J.}~\bibnamefont {Deng}}, \bibinfo {author} {\bibfnamefont
  {X.-B.}\ \bibnamefont {Liu}}, \bibinfo {author} {\bibfnamefont
  {X.}~\bibnamefont {Chen}}, \bibinfo {author} {\bibfnamefont {J.-G.}\
  \bibnamefont {Guo}},\ and\ \bibinfo {author} {\bibfnamefont {X.}~\bibnamefont
  {Chen}},\ }\bibfield  {title} {\bibinfo {title} {Highly robust reentrant
  superconductivity in csv3sb5 under pressure},\ }\href
  {https://doi.org/10.1088/0256-307x/38/5/057402} {\bibfield  {journal}
  {\bibinfo  {journal} {Chinese Physics Letters}\ }\textbf {\bibinfo {volume}
  {38}},\ \bibinfo {pages} {057402} (\bibinfo {year}
  {2021}{\natexlab{b}})}\BibitemShut {NoStop}%
\bibitem [{\citenamefont {Wen}\ \emph {et~al.}(2021)\citenamefont {Wen},
  \citenamefont {Zhu}, \citenamefont {Xiao}, \citenamefont {Hao}, \citenamefont
  {Mondaini}, \citenamefont {Guo},\ and\ \citenamefont
  {Feng}}]{wen2021superconducting}%
  \BibitemOpen
  \bibfield  {author} {\bibinfo {author} {\bibfnamefont {C.}~\bibnamefont
  {Wen}}, \bibinfo {author} {\bibfnamefont {X.}~\bibnamefont {Zhu}}, \bibinfo
  {author} {\bibfnamefont {Z.}~\bibnamefont {Xiao}}, \bibinfo {author}
  {\bibfnamefont {N.}~\bibnamefont {Hao}}, \bibinfo {author} {\bibfnamefont
  {R.}~\bibnamefont {Mondaini}}, \bibinfo {author} {\bibfnamefont
  {H.}~\bibnamefont {Guo}},\ and\ \bibinfo {author} {\bibfnamefont
  {S.}~\bibnamefont {Feng}},\ }\href@noop {} {\bibinfo {title} {superconducting
  pairing symmetry in the kagome-lattice hubbard model}} (\bibinfo {year}
  {2021}),\ \Eprint {https://arxiv.org/abs/2109.12582} {arXiv:2109.12582
  [cond-mat.str-el]} \BibitemShut {NoStop}%
\bibitem [{\citenamefont {Jiang}\ \emph {et~al.}(2021)\citenamefont {Jiang},
  \citenamefont {Yin}, \citenamefont {Denner}, \citenamefont {Shumiya},
  \citenamefont {Ortiz}, \citenamefont {Xu}, \citenamefont {Guguchia},
  \citenamefont {He}, \citenamefont {Hossain}, \citenamefont {Liu},
  \citenamefont {Ruff}, \citenamefont {Kautzsch}, \citenamefont {Zhang},
  \citenamefont {Chang}, \citenamefont {Belopolski}, \citenamefont {Zhang},
  \citenamefont {Cochran}, \citenamefont {Multer}, \citenamefont {Litskevich},
  \citenamefont {Cheng}, \citenamefont {Yang}, \citenamefont {Wang},
  \citenamefont {Thomale}, \citenamefont {Neupert}, \citenamefont {Wilson},\
  and\ \citenamefont {Hasan}}]{kagome_CDW1}%
  \BibitemOpen
  \bibfield  {author} {\bibinfo {author} {\bibfnamefont {Y.-X.}\ \bibnamefont
  {Jiang}}, \bibinfo {author} {\bibfnamefont {J.-X.}\ \bibnamefont {Yin}},
  \bibinfo {author} {\bibfnamefont {M.~M.}\ \bibnamefont {Denner}}, \bibinfo
  {author} {\bibfnamefont {N.}~\bibnamefont {Shumiya}}, \bibinfo {author}
  {\bibfnamefont {B.~R.}\ \bibnamefont {Ortiz}}, \bibinfo {author}
  {\bibfnamefont {G.}~\bibnamefont {Xu}}, \bibinfo {author} {\bibfnamefont
  {Z.}~\bibnamefont {Guguchia}}, \bibinfo {author} {\bibfnamefont
  {J.}~\bibnamefont {He}}, \bibinfo {author} {\bibfnamefont {M.~S.}\
  \bibnamefont {Hossain}}, \bibinfo {author} {\bibfnamefont {X.}~\bibnamefont
  {Liu}}, \bibinfo {author} {\bibfnamefont {J.}~\bibnamefont {Ruff}}, \bibinfo
  {author} {\bibfnamefont {L.}~\bibnamefont {Kautzsch}}, \bibinfo {author}
  {\bibfnamefont {S.~S.}\ \bibnamefont {Zhang}}, \bibinfo {author}
  {\bibfnamefont {G.}~\bibnamefont {Chang}}, \bibinfo {author} {\bibfnamefont
  {I.}~\bibnamefont {Belopolski}}, \bibinfo {author} {\bibfnamefont
  {Q.}~\bibnamefont {Zhang}}, \bibinfo {author} {\bibfnamefont {T.~A.}\
  \bibnamefont {Cochran}}, \bibinfo {author} {\bibfnamefont {D.}~\bibnamefont
  {Multer}}, \bibinfo {author} {\bibfnamefont {M.}~\bibnamefont {Litskevich}},
  \bibinfo {author} {\bibfnamefont {Z.-J.}\ \bibnamefont {Cheng}}, \bibinfo
  {author} {\bibfnamefont {X.~P.}\ \bibnamefont {Yang}}, \bibinfo {author}
  {\bibfnamefont {Z.}~\bibnamefont {Wang}}, \bibinfo {author} {\bibfnamefont
  {R.}~\bibnamefont {Thomale}}, \bibinfo {author} {\bibfnamefont
  {T.}~\bibnamefont {Neupert}}, \bibinfo {author} {\bibfnamefont {S.~D.}\
  \bibnamefont {Wilson}},\ and\ \bibinfo {author} {\bibfnamefont {M.~Z.}\
  \bibnamefont {Hasan}},\ }\bibfield  {title} {\bibinfo {title} {Unconventional
  chiral charge order in kagome superconductor kv3sb5},\ }\href
  {https://doi.org/10.1038/s41563-021-01034-y} {\bibfield  {journal} {\bibinfo
  {journal} {Nature Materials}\ }\textbf {\bibinfo {volume} {20}},\ \bibinfo
  {pages} {1353} (\bibinfo {year} {2021})}\BibitemShut {NoStop}%
\bibitem [{\citenamefont {Feng}\ \emph {et~al.}(2021)\citenamefont {Feng},
  \citenamefont {Jiang}, \citenamefont {Wang},\ and\ \citenamefont
  {Hu}}]{kagome_CDW2}%
  \BibitemOpen
  \bibfield  {author} {\bibinfo {author} {\bibfnamefont {X.}~\bibnamefont
  {Feng}}, \bibinfo {author} {\bibfnamefont {K.}~\bibnamefont {Jiang}},
  \bibinfo {author} {\bibfnamefont {Z.}~\bibnamefont {Wang}},\ and\ \bibinfo
  {author} {\bibfnamefont {J.}~\bibnamefont {Hu}},\ }\bibfield  {title}
  {\bibinfo {title} {Chiral flux phase in the kagome superconductor av3sb5},\
  }\href {https://doi.org/https://doi.org/10.1016/j.scib.2021.04.043}
  {\bibfield  {journal} {\bibinfo  {journal} {Science Bulletin}\ }\textbf
  {\bibinfo {volume} {66}},\ \bibinfo {pages} {1384} (\bibinfo {year}
  {2021})}\BibitemShut {NoStop}%
\bibitem [{\citenamefont {Denner}\ \emph {et~al.}(2021)\citenamefont {Denner},
  \citenamefont {Thomale},\ and\ \citenamefont {Neupert}}]{kagome_CDW3}%
  \BibitemOpen
  \bibfield  {author} {\bibinfo {author} {\bibfnamefont {M.~M.}\ \bibnamefont
  {Denner}}, \bibinfo {author} {\bibfnamefont {R.}~\bibnamefont {Thomale}},\
  and\ \bibinfo {author} {\bibfnamefont {T.}~\bibnamefont {Neupert}},\
  }\bibfield  {title} {\bibinfo {title} {Analysis of charge order in the kagome
  metal $a{\mathrm{v}}_{3}{\mathrm{sb}}_{5}$
  ($a=\mathrm{K},\mathrm{Rb},\mathrm{Cs}$)},\ }\href
  {https://doi.org/10.1103/PhysRevLett.127.217601} {\bibfield  {journal}
  {\bibinfo  {journal} {Physical Review Letters}\ }\textbf {\bibinfo {volume}
  {127}},\ \bibinfo {pages} {217601} (\bibinfo {year} {2021})}\BibitemShut
  {NoStop}%
\bibitem [{\citenamefont {Lin}\ and\ \citenamefont
  {Nandkishore}(2021{\natexlab{a}})}]{kagome_CDW4}%
  \BibitemOpen
  \bibfield  {author} {\bibinfo {author} {\bibfnamefont {Y.-P.}\ \bibnamefont
  {Lin}}\ and\ \bibinfo {author} {\bibfnamefont {R.~M.}\ \bibnamefont
  {Nandkishore}},\ }\bibfield  {title} {\bibinfo {title} {Complex charge
  density waves at van hove singularity on hexagonal lattices: Haldane-model
  phase diagram and potential realization in the kagome metals
  $a{V}_{3}{\mathrm{sb}}_{5}$ ($a$=k, rb, cs)},\ }\href
  {https://doi.org/10.1103/PhysRevB.104.045122} {\bibfield  {journal} {\bibinfo
   {journal} {Physical Review B}\ }\textbf {\bibinfo {volume} {104}},\ \bibinfo
  {pages} {045122} (\bibinfo {year} {2021}{\natexlab{a}})}\BibitemShut
  {NoStop}%
\bibitem [{\citenamefont {Neupert}\ \emph {et~al.}(2021)\citenamefont
  {Neupert}, \citenamefont {Denner}, \citenamefont {Yin}, \citenamefont
  {Thomale},\ and\ \citenamefont {Hasan}}]{chargeorder}%
  \BibitemOpen
  \bibfield  {author} {\bibinfo {author} {\bibfnamefont {T.}~\bibnamefont
  {Neupert}}, \bibinfo {author} {\bibfnamefont {M.~M.}\ \bibnamefont {Denner}},
  \bibinfo {author} {\bibfnamefont {J.-X.}\ \bibnamefont {Yin}}, \bibinfo
  {author} {\bibfnamefont {R.}~\bibnamefont {Thomale}},\ and\ \bibinfo {author}
  {\bibfnamefont {M.~Z.}\ \bibnamefont {Hasan}},\ }\bibfield  {title} {\bibinfo
  {title} {Charge order and superconductivity in kagome materials},\ }\bibfield
   {journal} {\bibinfo  {journal} {Nature Physics}\ }\href
  {https://doi.org/10.1038/s41567-021-01404-y} {10.1038/s41567-021-01404-y}
  (\bibinfo {year} {2021})\BibitemShut {NoStop}%
\bibitem [{\citenamefont {Lin}\ and\ \citenamefont
  {Nandkishore}(2021{\natexlab{b}})}]{lin2021kagome}%
  \BibitemOpen
  \bibfield  {author} {\bibinfo {author} {\bibfnamefont {Y.-P.}\ \bibnamefont
  {Lin}}\ and\ \bibinfo {author} {\bibfnamefont {R.~M.}\ \bibnamefont
  {Nandkishore}},\ }\href@noop {} {\bibinfo {title} {Kagome superconductors
  from pomeranchuk fluctuations in charge density wave metals}} (\bibinfo
  {year} {2021}{\natexlab{b}}),\ \Eprint {https://arxiv.org/abs/2107.09050}
  {arXiv:2107.09050 [cond-mat.str-el]} \BibitemShut {NoStop}%
\bibitem [{\citenamefont {Li}\ \emph {et~al.}(2021)\citenamefont {Li},
  \citenamefont {Zhang}, \citenamefont {Yilmaz}, \citenamefont {Pai},
  \citenamefont {Marvinney}, \citenamefont {Said}, \citenamefont {Yin},
  \citenamefont {Gong}, \citenamefont {Tu}, \citenamefont {Vescovo},
  \citenamefont {Nelson}, \citenamefont {Moore}, \citenamefont {Murakami},
  \citenamefont {Lei}, \citenamefont {Lee}, \citenamefont {Lawrie},\ and\
  \citenamefont {Miao}}]{x-ray}%
  \BibitemOpen
  \bibfield  {author} {\bibinfo {author} {\bibfnamefont {H.}~\bibnamefont
  {Li}}, \bibinfo {author} {\bibfnamefont {T.}~\bibnamefont {Zhang}}, \bibinfo
  {author} {\bibfnamefont {T.}~\bibnamefont {Yilmaz}}, \bibinfo {author}
  {\bibfnamefont {Y.}~\bibnamefont {Pai}}, \bibinfo {author} {\bibfnamefont
  {C.}~\bibnamefont {Marvinney}}, \bibinfo {author} {\bibfnamefont
  {A.}~\bibnamefont {Said}}, \bibinfo {author} {\bibfnamefont {Q.}~\bibnamefont
  {Yin}}, \bibinfo {author} {\bibfnamefont {C.}~\bibnamefont {Gong}}, \bibinfo
  {author} {\bibfnamefont {Z.}~\bibnamefont {Tu}}, \bibinfo {author}
  {\bibfnamefont {E.}~\bibnamefont {Vescovo}}, \bibinfo {author} {\bibfnamefont
  {C.}~\bibnamefont {Nelson}}, \bibinfo {author} {\bibfnamefont
  {R.}~\bibnamefont {Moore}}, \bibinfo {author} {\bibfnamefont
  {S.}~\bibnamefont {Murakami}}, \bibinfo {author} {\bibfnamefont
  {H.}~\bibnamefont {Lei}}, \bibinfo {author} {\bibfnamefont {H.}~\bibnamefont
  {Lee}}, \bibinfo {author} {\bibfnamefont {B.}~\bibnamefont {Lawrie}},\ and\
  \bibinfo {author} {\bibfnamefont {H.}~\bibnamefont {Miao}},\ }\bibfield
  {title} {\bibinfo {title} {Observation of unconventional charge density wave
  without acoustic phonon anomaly in kagome superconductors av3sb5 (a=rb,
  cs)},\ }\href {https://doi.org/10.1103/PhysRevX.11.031050} {\bibfield
  {journal} {\bibinfo  {journal} {Physical Review X}\ }\textbf {\bibinfo
  {volume} {11}},\ \bibinfo {pages} {031050} (\bibinfo {year}
  {2021})}\BibitemShut {NoStop}%
\bibitem [{\citenamefont {Duan}\ \emph {et~al.}(2021)\citenamefont {Duan},
  \citenamefont {Nie}, \citenamefont {Luo}, \citenamefont {Yu}, \citenamefont
  {Ortiz}, \citenamefont {Yin}, \citenamefont {Su}, \citenamefont {Du},
  \citenamefont {Wang}, \citenamefont {Chen}, \citenamefont {Lu}, \citenamefont
  {Ying}, \citenamefont {Wilson}, \citenamefont {Chen}, \citenamefont {Song},\
  and\ \citenamefont {Yuan}}]{Duan2021}%
  \BibitemOpen
  \bibfield  {author} {\bibinfo {author} {\bibfnamefont {W.}~\bibnamefont
  {Duan}}, \bibinfo {author} {\bibfnamefont {Z.}~\bibnamefont {Nie}}, \bibinfo
  {author} {\bibfnamefont {S.}~\bibnamefont {Luo}}, \bibinfo {author}
  {\bibfnamefont {F.}~\bibnamefont {Yu}}, \bibinfo {author} {\bibfnamefont
  {B.~R.}\ \bibnamefont {Ortiz}}, \bibinfo {author} {\bibfnamefont
  {L.}~\bibnamefont {Yin}}, \bibinfo {author} {\bibfnamefont {H.}~\bibnamefont
  {Su}}, \bibinfo {author} {\bibfnamefont {F.}~\bibnamefont {Du}}, \bibinfo
  {author} {\bibfnamefont {A.}~\bibnamefont {Wang}}, \bibinfo {author}
  {\bibfnamefont {Y.}~\bibnamefont {Chen}}, \bibinfo {author} {\bibfnamefont
  {X.}~\bibnamefont {Lu}}, \bibinfo {author} {\bibfnamefont {J.}~\bibnamefont
  {Ying}}, \bibinfo {author} {\bibfnamefont {S.~D.}\ \bibnamefont {Wilson}},
  \bibinfo {author} {\bibfnamefont {X.}~\bibnamefont {Chen}}, \bibinfo {author}
  {\bibfnamefont {Y.}~\bibnamefont {Song}},\ and\ \bibinfo {author}
  {\bibfnamefont {H.}~\bibnamefont {Yuan}},\ }\bibfield  {title} {\bibinfo
  {title} {Nodeless superconductivity in the kagome metal csv3sb5},\ }\href
  {https://doi.org/10.1007/s11433-021-1747-7} {\bibfield  {journal} {\bibinfo
  {journal} {Science China Physics, Mechanics, and Astronomy}\ }\textbf
  {\bibinfo {volume} {64}},\ \bibinfo {pages} {107462} (\bibinfo {year}
  {2021})}\BibitemShut {NoStop}%
\bibitem [{\citenamefont {Chen}\ \emph
  {et~al.}(2021{\natexlab{c}})\citenamefont {Chen}, \citenamefont {Yang},
  \citenamefont {Hu}, \citenamefont {Zhao}, \citenamefont {Yuan}, \citenamefont
  {Xing}, \citenamefont {Qian}, \citenamefont {Huang}, \citenamefont {Li},
  \citenamefont {Ye}, \citenamefont {Ma}, \citenamefont {Ni}, \citenamefont
  {Zhang}, \citenamefont {Yin}, \citenamefont {Gong}, \citenamefont {Tu},
  \citenamefont {Lei}, \citenamefont {Tan}, \citenamefont {Zhou}, \citenamefont
  {Shen}, \citenamefont {Dong}, \citenamefont {Yan}, \citenamefont {Wang},\
  and\ \citenamefont {Gao}}]{Chen2021a}%
  \BibitemOpen
  \bibfield  {author} {\bibinfo {author} {\bibfnamefont {H.}~\bibnamefont
  {Chen}}, \bibinfo {author} {\bibfnamefont {H.}~\bibnamefont {Yang}}, \bibinfo
  {author} {\bibfnamefont {B.}~\bibnamefont {Hu}}, \bibinfo {author}
  {\bibfnamefont {Z.}~\bibnamefont {Zhao}}, \bibinfo {author} {\bibfnamefont
  {J.}~\bibnamefont {Yuan}}, \bibinfo {author} {\bibfnamefont {Y.}~\bibnamefont
  {Xing}}, \bibinfo {author} {\bibfnamefont {G.}~\bibnamefont {Qian}}, \bibinfo
  {author} {\bibfnamefont {Z.}~\bibnamefont {Huang}}, \bibinfo {author}
  {\bibfnamefont {G.}~\bibnamefont {Li}}, \bibinfo {author} {\bibfnamefont
  {Y.}~\bibnamefont {Ye}}, \bibinfo {author} {\bibfnamefont {S.}~\bibnamefont
  {Ma}}, \bibinfo {author} {\bibfnamefont {S.}~\bibnamefont {Ni}}, \bibinfo
  {author} {\bibfnamefont {H.}~\bibnamefont {Zhang}}, \bibinfo {author}
  {\bibfnamefont {Q.}~\bibnamefont {Yin}}, \bibinfo {author} {\bibfnamefont
  {C.}~\bibnamefont {Gong}}, \bibinfo {author} {\bibfnamefont {Z.}~\bibnamefont
  {Tu}}, \bibinfo {author} {\bibfnamefont {H.}~\bibnamefont {Lei}}, \bibinfo
  {author} {\bibfnamefont {H.}~\bibnamefont {Tan}}, \bibinfo {author}
  {\bibfnamefont {S.}~\bibnamefont {Zhou}}, \bibinfo {author} {\bibfnamefont
  {C.}~\bibnamefont {Shen}}, \bibinfo {author} {\bibfnamefont {X.}~\bibnamefont
  {Dong}}, \bibinfo {author} {\bibfnamefont {B.}~\bibnamefont {Yan}}, \bibinfo
  {author} {\bibfnamefont {Z.}~\bibnamefont {Wang}},\ and\ \bibinfo {author}
  {\bibfnamefont {H.-J.}\ \bibnamefont {Gao}},\ }\href
  {https://ui.adsabs.harvard.edu/abs/2021arXiv210309188C} {\bibinfo {title}
  {Roton pair density wave and unconventional strong-coupling superconductivity
  in a topological kagome metal}} (\bibinfo {year}
  {2021}{\natexlab{c}})\BibitemShut {NoStop}%
\bibitem [{\citenamefont {Zhao}\ \emph
  {et~al.}(2021{\natexlab{b}})\citenamefont {Zhao}, \citenamefont {Wang},
  \citenamefont {Xia}, \citenamefont {Yin}, \citenamefont {Ni}, \citenamefont
  {Huang}, \citenamefont {Tu}, \citenamefont {Tao}, \citenamefont {Tu},
  \citenamefont {Gong}, \citenamefont {Lei}, \citenamefont {Guo}, \citenamefont
  {Yang},\ and\ \citenamefont {Li}}]{zhao2021nodal}%
  \BibitemOpen
  \bibfield  {author} {\bibinfo {author} {\bibfnamefont {C.~C.}\ \bibnamefont
  {Zhao}}, \bibinfo {author} {\bibfnamefont {L.~S.}\ \bibnamefont {Wang}},
  \bibinfo {author} {\bibfnamefont {W.}~\bibnamefont {Xia}}, \bibinfo {author}
  {\bibfnamefont {Q.~W.}\ \bibnamefont {Yin}}, \bibinfo {author} {\bibfnamefont
  {J.~M.}\ \bibnamefont {Ni}}, \bibinfo {author} {\bibfnamefont {Y.~Y.}\
  \bibnamefont {Huang}}, \bibinfo {author} {\bibfnamefont {C.~P.}\ \bibnamefont
  {Tu}}, \bibinfo {author} {\bibfnamefont {Z.~C.}\ \bibnamefont {Tao}},
  \bibinfo {author} {\bibfnamefont {Z.~J.}\ \bibnamefont {Tu}}, \bibinfo
  {author} {\bibfnamefont {C.~S.}\ \bibnamefont {Gong}}, \bibinfo {author}
  {\bibfnamefont {H.~C.}\ \bibnamefont {Lei}}, \bibinfo {author} {\bibfnamefont
  {Y.~F.}\ \bibnamefont {Guo}}, \bibinfo {author} {\bibfnamefont {X.~F.}\
  \bibnamefont {Yang}},\ and\ \bibinfo {author} {\bibfnamefont {S.~Y.}\
  \bibnamefont {Li}},\ }\href
  {https://ui.adsabs.harvard.edu/abs/2021arXiv210208356Z} {\bibinfo {title}
  {Nodal superconductivity and superconducting domes in the topological kagome
  metal csv3sb5}} (\bibinfo {year} {2021}{\natexlab{b}})\BibitemShut {NoStop}%
\bibitem [{\citenamefont {au2}\ \emph {et~al.}(2021)\citenamefont {au2},
  \citenamefont {Das}, \citenamefont {Yin}, \citenamefont {Liu}, \citenamefont
  {Gupta}, \citenamefont {Jiang}, \citenamefont {Medarde}, \citenamefont {Wu},
  \citenamefont {Lei}, \citenamefont {Chang}, \citenamefont {Dai},
  \citenamefont {Si}, \citenamefont {Miao}, \citenamefont {Thomale},
  \citenamefont {Neupert}, \citenamefont {Shi}, \citenamefont {Khasanov},
  \citenamefont {Hasan}, \citenamefont {Luetkens},\ and\ \citenamefont
  {Guguchia}}]{mielke2021timereversal}%
  \BibitemOpen
  \bibfield  {author} {\bibinfo {author} {\bibfnamefont {C.~M.~I.}\
  \bibnamefont {au2}}, \bibinfo {author} {\bibfnamefont {D.}~\bibnamefont
  {Das}}, \bibinfo {author} {\bibfnamefont {J.~X.}\ \bibnamefont {Yin}},
  \bibinfo {author} {\bibfnamefont {H.}~\bibnamefont {Liu}}, \bibinfo {author}
  {\bibfnamefont {R.}~\bibnamefont {Gupta}}, \bibinfo {author} {\bibfnamefont
  {Y.~X.}\ \bibnamefont {Jiang}}, \bibinfo {author} {\bibfnamefont
  {M.}~\bibnamefont {Medarde}}, \bibinfo {author} {\bibfnamefont
  {X.}~\bibnamefont {Wu}}, \bibinfo {author} {\bibfnamefont {H.~C.}\
  \bibnamefont {Lei}}, \bibinfo {author} {\bibfnamefont {J.~J.}\ \bibnamefont
  {Chang}}, \bibinfo {author} {\bibfnamefont {P.}~\bibnamefont {Dai}}, \bibinfo
  {author} {\bibfnamefont {Q.}~\bibnamefont {Si}}, \bibinfo {author}
  {\bibfnamefont {H.}~\bibnamefont {Miao}}, \bibinfo {author} {\bibfnamefont
  {R.}~\bibnamefont {Thomale}}, \bibinfo {author} {\bibfnamefont
  {T.}~\bibnamefont {Neupert}}, \bibinfo {author} {\bibfnamefont
  {Y.}~\bibnamefont {Shi}}, \bibinfo {author} {\bibfnamefont {R.}~\bibnamefont
  {Khasanov}}, \bibinfo {author} {\bibfnamefont {M.~Z.}\ \bibnamefont {Hasan}},
  \bibinfo {author} {\bibfnamefont {H.}~\bibnamefont {Luetkens}},\ and\
  \bibinfo {author} {\bibfnamefont {Z.}~\bibnamefont {Guguchia}},\ }\href@noop
  {} {\bibinfo {title} {Time-reversal symmetry-breaking charge order in a
  kagome superconductor}} (\bibinfo {year} {2021}),\ \Eprint
  {https://arxiv.org/abs/2106.13443} {arXiv:2106.13443 [cond-mat.mtrl-sci]}
  \BibitemShut {NoStop}%
\bibitem [{\citenamefont {Liang}\ \emph {et~al.}(2021)\citenamefont {Liang},
  \citenamefont {Hou}, \citenamefont {Zhang}, \citenamefont {Ma}, \citenamefont
  {Wu}, \citenamefont {Zhang}, \citenamefont {Yu}, \citenamefont {Ying},
  \citenamefont {Jiang}, \citenamefont {Shan}, \citenamefont {Wang},\ and\
  \citenamefont {Chen}}]{vortex_STM1}%
  \BibitemOpen
  \bibfield  {author} {\bibinfo {author} {\bibfnamefont {Z.}~\bibnamefont
  {Liang}}, \bibinfo {author} {\bibfnamefont {X.}~\bibnamefont {Hou}}, \bibinfo
  {author} {\bibfnamefont {F.}~\bibnamefont {Zhang}}, \bibinfo {author}
  {\bibfnamefont {W.}~\bibnamefont {Ma}}, \bibinfo {author} {\bibfnamefont
  {P.}~\bibnamefont {Wu}}, \bibinfo {author} {\bibfnamefont {Z.}~\bibnamefont
  {Zhang}}, \bibinfo {author} {\bibfnamefont {F.}~\bibnamefont {Yu}}, \bibinfo
  {author} {\bibfnamefont {J.~J.}\ \bibnamefont {Ying}}, \bibinfo {author}
  {\bibfnamefont {K.}~\bibnamefont {Jiang}}, \bibinfo {author} {\bibfnamefont
  {L.}~\bibnamefont {Shan}}, \bibinfo {author} {\bibfnamefont {Z.}~\bibnamefont
  {Wang}},\ and\ \bibinfo {author} {\bibfnamefont {X.~H.}\ \bibnamefont
  {Chen}},\ }\bibfield  {title} {\bibinfo {title} {Three-dimensional charge
  density wave and surface-dependent vortex-core states in a kagome
  superconductor ${\mathrm{csv}}_{3}{\mathrm{sb}}_{5}$},\ }\href
  {https://doi.org/10.1103/PhysRevX.11.031026} {\bibfield  {journal} {\bibinfo
  {journal} {Physical Review X}\ }\textbf {\bibinfo {volume} {11}},\ \bibinfo
  {pages} {031026} (\bibinfo {year} {2021})}\BibitemShut {NoStop}%
\bibitem [{\citenamefont {Xu}\ \emph {et~al.}(2021)\citenamefont {Xu},
  \citenamefont {Yan}, \citenamefont {Yin}, \citenamefont {Xia}, \citenamefont
  {Fang}, \citenamefont {Chen}, \citenamefont {Li}, \citenamefont {Yang},
  \citenamefont {Guo},\ and\ \citenamefont {Feng}}]{vortex_STM2}%
  \BibitemOpen
  \bibfield  {author} {\bibinfo {author} {\bibfnamefont {H.-S.}\ \bibnamefont
  {Xu}}, \bibinfo {author} {\bibfnamefont {Y.-J.}\ \bibnamefont {Yan}},
  \bibinfo {author} {\bibfnamefont {R.}~\bibnamefont {Yin}}, \bibinfo {author}
  {\bibfnamefont {W.}~\bibnamefont {Xia}}, \bibinfo {author} {\bibfnamefont
  {S.}~\bibnamefont {Fang}}, \bibinfo {author} {\bibfnamefont {Z.}~\bibnamefont
  {Chen}}, \bibinfo {author} {\bibfnamefont {Y.}~\bibnamefont {Li}}, \bibinfo
  {author} {\bibfnamefont {W.}~\bibnamefont {Yang}}, \bibinfo {author}
  {\bibfnamefont {Y.}~\bibnamefont {Guo}},\ and\ \bibinfo {author}
  {\bibfnamefont {D.-L.}\ \bibnamefont {Feng}},\ }\bibfield  {title} {\bibinfo
  {title} {Multiband superconductivity with sign-preserving order parameter in
  kagome superconductor ${\mathrm{csv}}_{3}{\mathrm{sb}}_{5}$},\ }\href
  {https://doi.org/10.1103/PhysRevLett.127.187004} {\bibfield  {journal}
  {\bibinfo  {journal} {Physical Review Letters}\ }\textbf {\bibinfo {volume}
  {127}},\ \bibinfo {pages} {187004} (\bibinfo {year} {2021})}\BibitemShut
  {NoStop}%
\bibitem [{\citenamefont {Wu}\ \emph {et~al.}(2021)\citenamefont {Wu},
  \citenamefont {Schwemmer}, \citenamefont {Müller}, \citenamefont
  {Consiglio}, \citenamefont {Sangiovanni}, \citenamefont {Di~Sante},
  \citenamefont {Iqbal}, \citenamefont {Hanke}, \citenamefont {Schnyder},
  \citenamefont {Denner}, \citenamefont {Fischer}, \citenamefont {Neupert},\
  and\ \citenamefont {Thomale}}]{RPA}%
  \BibitemOpen
  \bibfield  {author} {\bibinfo {author} {\bibfnamefont {X.}~\bibnamefont
  {Wu}}, \bibinfo {author} {\bibfnamefont {T.}~\bibnamefont {Schwemmer}},
  \bibinfo {author} {\bibfnamefont {T.}~\bibnamefont {Müller}}, \bibinfo
  {author} {\bibfnamefont {A.}~\bibnamefont {Consiglio}}, \bibinfo {author}
  {\bibfnamefont {G.}~\bibnamefont {Sangiovanni}}, \bibinfo {author}
  {\bibfnamefont {D.}~\bibnamefont {Di~Sante}}, \bibinfo {author}
  {\bibfnamefont {Y.}~\bibnamefont {Iqbal}}, \bibinfo {author} {\bibfnamefont
  {W.}~\bibnamefont {Hanke}}, \bibinfo {author} {\bibfnamefont {A.~P.}\
  \bibnamefont {Schnyder}}, \bibinfo {author} {\bibfnamefont {M.~M.}\
  \bibnamefont {Denner}}, \bibinfo {author} {\bibfnamefont {M.~H.}\
  \bibnamefont {Fischer}}, \bibinfo {author} {\bibfnamefont {T.}~\bibnamefont
  {Neupert}},\ and\ \bibinfo {author} {\bibfnamefont {R.}~\bibnamefont
  {Thomale}},\ }\bibfield  {title} {\bibinfo {title} {Nature of unconventional
  pairing in the kagome superconductors $a{\mathrm{v}}_{3}{\mathrm{sb}}_{5}$
  ($a=\mathrm{K},\mathrm{Rb},\mathrm{Cs}$)},\ }\href
  {https://doi.org/10.1103/PhysRevLett.127.177001} {\bibfield  {journal}
  {\bibinfo  {journal} {Physical Review Letters}\ }\textbf {\bibinfo {volume}
  {127}},\ \bibinfo {pages} {177001} (\bibinfo {year} {2021})}\BibitemShut
  {NoStop}%
\bibitem [{\citenamefont {Han}(2007)}]{f-wave}%
  \BibitemOpen
  \bibfield  {author} {\bibinfo {author} {\bibfnamefont {Q.}~\bibnamefont
  {Han}},\ }\bibfield  {title} {\bibinfo {title} {Vortex state in f-wave
  superconductors},\ }\href {https://doi.org/10.1142/s0217984907013377}
  {\bibfield  {journal} {\bibinfo  {journal} {Modern Physics Letters B}\
  }\textbf {\bibinfo {volume} {21}},\ \bibinfo {pages} {1051} (\bibinfo {year}
  {2007})}\BibitemShut {NoStop}%
\bibitem [{\citenamefont {Caroli}\ \emph {et~al.}(1964)\citenamefont {Caroli},
  \citenamefont {De~Gennes},\ and\ \citenamefont {Matricon}}]{deGennes}%
  \BibitemOpen
  \bibfield  {author} {\bibinfo {author} {\bibfnamefont {C.}~\bibnamefont
  {Caroli}}, \bibinfo {author} {\bibfnamefont {P.~G.}\ \bibnamefont
  {De~Gennes}},\ and\ \bibinfo {author} {\bibfnamefont {J.}~\bibnamefont
  {Matricon}},\ }\bibfield  {title} {\bibinfo {title} {Bound fermion states on
  a vortex line in a type ii superconductor},\ }\href
  {https://doi.org/https://doi.org/10.1016/0031-9163(64)90375-0} {\bibfield
  {journal} {\bibinfo  {journal} {Physics Letters}\ }\textbf {\bibinfo {volume}
  {9}},\ \bibinfo {pages} {307} (\bibinfo {year} {1964})}\BibitemShut {NoStop}%
\bibitem [{\citenamefont {Volovik}(1999)}]{Volovik}%
  \BibitemOpen
  \bibfield  {author} {\bibinfo {author} {\bibfnamefont {G.~E.}\ \bibnamefont
  {Volovik}},\ }\bibfield  {title} {\bibinfo {title} {Fermion zero modes on
  vortices in chiral superconductors},\ }\href
  {https://doi.org/10.1134/1.568223} {\bibfield  {journal} {\bibinfo  {journal}
  {Journal of Experimental and Theoretical Physics Letters}\ }\textbf {\bibinfo
  {volume} {70}},\ \bibinfo {pages} {609} (\bibinfo {year} {1999})}\BibitemShut
  {NoStop}%
\bibitem [{\citenamefont {Zhu}(2016)}]{BdGmethod}%
  \BibitemOpen
  \bibfield  {author} {\bibinfo {author} {\bibfnamefont {J.-X.}\ \bibnamefont
  {Zhu}},\ }\bibfield  {title} {\bibinfo {title} {\textit{Bogoliubov-de Gennes
  Method and its Applications}},\ }\href
  {https://link.springer.com/book/10.1007/978-3-319-31314-6} {\bibfield
  {journal} {\bibinfo  {journal} {Springer, New York}\ } (\bibinfo {year}
  {2016})}\BibitemShut {NoStop}%
\bibitem [{\citenamefont {Hayashi}\ \emph {et~al.}(1998)\citenamefont
  {Hayashi}, \citenamefont {Isoshima}, \citenamefont {Ichioka},\ and\
  \citenamefont {Machida}}]{low_lying}%
  \BibitemOpen
  \bibfield  {author} {\bibinfo {author} {\bibfnamefont {N.}~\bibnamefont
  {Hayashi}}, \bibinfo {author} {\bibfnamefont {T.}~\bibnamefont {Isoshima}},
  \bibinfo {author} {\bibfnamefont {M.}~\bibnamefont {Ichioka}},\ and\ \bibinfo
  {author} {\bibfnamefont {K.}~\bibnamefont {Machida}},\ }\bibfield  {title}
  {\bibinfo {title} {Low-lying quasiparticle excitations around a vortex core
  in quantum limit},\ }\href {https://doi.org/10.1103/PhysRevLett.80.2921}
  {\bibfield  {journal} {\bibinfo  {journal} {Physical Review Letters}\
  }\textbf {\bibinfo {volume} {80}},\ \bibinfo {pages} {2921} (\bibinfo {year}
  {1998})}\BibitemShut {NoStop}%
\bibitem [{\citenamefont {Alicea}(2012)}]{review_p_wave}%
  \BibitemOpen
  \bibfield  {author} {\bibinfo {author} {\bibfnamefont {J.}~\bibnamefont
  {Alicea}},\ }\bibfield  {title} {\bibinfo {title} {New directions in the
  pursuit of majorana fermions in solid state systems},\ }\href
  {https://doi.org/10.1088/0034-4885/75/7/076501} {\bibfield  {journal}
  {\bibinfo  {journal} {Reports on Progress in Physics}\ }\textbf {\bibinfo
  {volume} {75}},\ \bibinfo {pages} {076501} (\bibinfo {year}
  {2012})}\BibitemShut {NoStop}%
\bibitem [{\citenamefont {Gygi}\ and\ \citenamefont
  {Schlüter}(1991)}]{chiral_d}%
  \BibitemOpen
  \bibfield  {author} {\bibinfo {author} {\bibfnamefont {F.}~\bibnamefont
  {Gygi}}\ and\ \bibinfo {author} {\bibfnamefont {M.}~\bibnamefont
  {Schlüter}},\ }\bibfield  {title} {\bibinfo {title} {Self-consistent
  electronic structure of a vortex line in a type-ii superconductor},\ }\href
  {https://doi.org/10.1103/PhysRevB.43.7609} {\bibfield  {journal} {\bibinfo
  {journal} {Physical Review B}\ }\textbf {\bibinfo {volume} {43}},\ \bibinfo
  {pages} {7609} (\bibinfo {year} {1991})}\BibitemShut {NoStop}%
\bibitem [{\citenamefont {Chern}(2016)}]{d-wave_topology}%
  \BibitemOpen
  \bibfield  {author} {\bibinfo {author} {\bibfnamefont {T.}~\bibnamefont
  {Chern}},\ }\bibfield  {title} {\bibinfo {title} {d + id and d wave
  topological superconductors and new mechanisms for bulk boundary
  correspondence},\ }\href {https://doi.org/10.1063/1.4961462} {\bibfield
  {journal} {\bibinfo  {journal} {AIP Advances}\ }\textbf {\bibinfo {volume}
  {6}},\ \bibinfo {pages} {085211} (\bibinfo {year} {2016})}\BibitemShut
  {NoStop}%
\bibitem [{\citenamefont {Sato}\ \emph {et~al.}(2011)\citenamefont {Sato},
  \citenamefont {Tanaka}, \citenamefont {Yada},\ and\ \citenamefont
  {Yokoyama}}]{Sato2011}%
  \BibitemOpen
  \bibfield  {author} {\bibinfo {author} {\bibfnamefont {M.}~\bibnamefont
  {Sato}}, \bibinfo {author} {\bibfnamefont {Y.}~\bibnamefont {Tanaka}},
  \bibinfo {author} {\bibfnamefont {K.}~\bibnamefont {Yada}},\ and\ \bibinfo
  {author} {\bibfnamefont {T.}~\bibnamefont {Yokoyama}},\ }\bibfield  {title}
  {\bibinfo {title} {Topology of andreev bound states with flat dispersion},\
  }\href {https://doi.org/10.1103/PhysRevB.83.224511} {\bibfield  {journal}
  {\bibinfo  {journal} {Physical Review B}\ }\textbf {\bibinfo {volume} {83}},\
  \bibinfo {pages} {224511} (\bibinfo {year} {2011})}\BibitemShut {NoStop}%
\bibitem [{\citenamefont {Schnyder}\ \emph {et~al.}(2012)\citenamefont
  {Schnyder}, \citenamefont {Brydon},\ and\ \citenamefont
  {Timm}}]{Schnyder2012}%
  \BibitemOpen
  \bibfield  {author} {\bibinfo {author} {\bibfnamefont {A.~P.}\ \bibnamefont
  {Schnyder}}, \bibinfo {author} {\bibfnamefont {P.~M.~R.}\ \bibnamefont
  {Brydon}},\ and\ \bibinfo {author} {\bibfnamefont {C.}~\bibnamefont {Timm}},\
  }\bibfield  {title} {\bibinfo {title} {Types of topological surface states in
  nodal noncentrosymmetric superconductors},\ }\href
  {https://doi.org/10.1103/PhysRevB.85.024522} {\bibfield  {journal} {\bibinfo
  {journal} {Physical Review B}\ }\textbf {\bibinfo {volume} {85}},\ \bibinfo
  {pages} {024522} (\bibinfo {year} {2012})}\BibitemShut {NoStop}%
\end{thebibliography}%

\end{document}